\documentclass[11pt,preprintnumbers,superscriptaddress,amsmath,amssymb,nofootinbib]{revtex4}

\newcommand{\PI}{\textrm{PI}}

\usepackage{graphicx}
\usepackage{dcolumn}
\usepackage{bm}
\usepackage{amssymb}
\usepackage{amsmath}
\usepackage{epsfig}    
\usepackage{color}
\usepackage{slashed}
\usepackage[hypertex]{hyperref}
\begin{document} 

\title{One-loop corrections to the Higgs self-couplings in the singlet extension}

\author{Shinya Kanemura}
\email{kanemu@sci.u-toyama.ac.jp}
\affiliation{Department of Physics, University of Toyama, \\3190 Gofuku, Toyama 930-8555, Japan}
\author{Mariko Kikuchi\footnote{Address after August 2016, National Taiwan University. }}
\email{kikuchi@jodo.sci.u-toyama.ac.jp}
\affiliation{Department of Physics, University of Toyama, \\3190 Gofuku, Toyama 930-8555, Japan}
 \author{Kei Yagyu}
\email{k.yagyu@soton.ac.uk}
\affiliation{School of Physics and Astronomy, University of Southampton, Southampton, SO17 1BJ, United Kingdom}
\preprint{UT-HET 116}
\begin{abstract}
We investigate predictions on the triple Higgs boson couplings with radiative corrections 
in the model with an additional real singlet scalar field.  
In this model, the second physical scalar state ($H$) appears in addition to 
the Higgs boson ($h$) with the mass 125 GeV. 
The $hhh$ vertex  is calculated at the one-loop level, 
and its possible deviation from the predictions in the standard model is 
evaluated under various theoretical constraints.
The decay rate of $H \to hh$ is also computed at the one-loop level. 
We also take into account the bound from the precise measurement of the $W$ boson mass, 
which gives the upper limit on the mixing angle $\alpha$ between 
two physical Higgs bosons for a given value of the mass of $H$ ($m_H^{}$). 
We find that the deviation in the $hhh$ coupling from the prediction in the standard model can maximally 
be about 250\%, 150\% and 75\% for $m_H^{}=300$, 500 and 1000 GeV, respectively,  
under the requirement that the cutoff scale of the model is higher than 3 TeV.    
We also discuss deviations from the standard model prediction 
in double Higgs boson production from the gluon fusion at the LHC
using the one-loop corrected Higgs boson vertices.

\end{abstract}
\maketitle

\section{Introduction}\label{sec:int}

Although the Higgs boson was found and its properties turned out to be consistent with the Standard Model (SM) for particle physics, 
we still do not know the Higgs sector, in particular, the structure of the Higgs potential and physics behind the electroweak symmetry breaking. 
On the other hand, there are several phenomena which cannot be explained in the SM such as neutrino oscillations, dark matter, baryon asymmetry of the Universe 
and cosmic inflation, which provide strong motivations to construct new models beyond the SM. 
If the origins of these phenomena are in the physics at the TeV scale, they are expected to be related to the physics of the Higgs sector. 
In such a case, the Higgs sector takes an extended form from the minimal model with an isospin doublet scalar field. 

In general, the non-minimal shape of the Higgs sector affects various observables.  
In particular, it gives deviations in the couplings of the discovered Higgs boson with the mass 125 GeV. 
Although there is no significant anomaly found in the current LHC data, 
the deviations might be detected in the future when the data will be more accumulated. 
Once the deviation is found, we may be able to obtain important information about the physics beyond the SM by fingerprinting 
the pattern of the deviation in various Higgs observables and the predictions in many new physics models~\cite{finger-tree}.  
In addition to many analyses at the tree level, radiative corrections to the Higgs boson couplings 
are evaluated in various extended Higgs sectors: e.g., two Higgs doublet models (THDMs)~\cite{THDM1,THDM2,THDM3,KOSY,Arhrib,Santos,Sakurai}, 
models with a singlet scalar field~\cite{HSM_KKY,Robens_reno,Costa,Camargo} and those with a triplet scalar field~\cite{HTM1,HTM2}, 
and new physics models: e.g., the minimal supersymmetric SM~\cite{MSSM1,MSSM2,MSSM3,MSSM4,MSSM5,MSSM6,Wu} and the minimal composite Higgs models~\cite{MCHM1,MCHM2,MCHM3}
in order to compare the theory predictions to the 
future precision data at High Luminosity LHC and future lepton colliders such as the International Linear Collider (ILC)~\cite{ILC,ILC2,ILC3}, 
the Compact LInear Collider (CLIC)~\cite{CLIC} and the Future $e^+e^-$ Circular Collider (FCCee). 

However, in order to obtain direct information on the Higgs potential, the measurement of the triple Higgs boson coupling is  
inevitable, which is one of the most important tasks of future collider experiments. 
From the information of the Higgs potential, we can approach to the physics behind electroweak symmetry breaking. 
It is known that in extended Higgs sectors physics predicting strongly first order phase transition 
simultaneously predicts a significant deviation in the triple Higgs boson coupling~\cite{1opt1,1opt2}.
In Ref.~\cite{Matsui}, synergy between measurements of gravitational waves and the triple Higgs boson coupling is discussed 
in probing the first-order electroweak phase transition. 
Therefore, the measurement of the triple Higgs boson coupling
is important not only to test the dynamics of electroweak symmetry breaking but also 
to investigate physics of the electroweak phase transition and scenarios of electroweak baryogenesis~\cite{EWBG1,EWBG2,EWBG3,1opt3}. 

In this paper, we focus on the Higgs Singlet Model (HSM) whose Higgs sector is composed of an isospin complex doublet field and a real singlet scalar field. 
The HSM has been drawn much attention in various interests in many papers. 
For example, works related to the electroweak baryogenesis have been done in Refs.~\cite{HSM-EWBG1,HSM-EWBG2,HSM-EWBG3,HSM-EWBG4,HSM-EWBG5,Fuyuto}.
Singlet scalar fields have also been studied in the context of the Higgs portal dark matter scenario~\cite{HSM-DM1,HSM-DM2,HSM-DM3,HSM-DM4}. 
The collider phenomenology, especially on the double Higgs boson production process at the LHC $gg\to hh$, has been calculated 
at the leading order (LO) in Ref.~\cite{CDL} and the next-to-leading order (NLO) in QCD in Ref.~\cite{DL} . 
Bounds on the parameter space in the HSM have been comprehensively surveyed by using data at the LHC Run-I in Ref.~\cite{Robens}. 

In addition to the above studies, 
there are papers for electroweak radiative corrections to the Higgs boson couplings in the HSM. 
In Ref.~\cite{HSM_KKY}, the $h$ couplings with weak bosons and fermions have been calculated at the one-loop level. 
In Ref.~\cite{Robens_reno}, one-loop corrections to the decay rate of the $H\to hh$ process with $H$ being a heavier Higgs boson. 
In this paper, we investigate one-loop corrections to the triple scalar boson couplings $hhh$ and $Hhh$
based on the on-shell renormalization scheme. 
We apply these one-loop corrected vertices to calculate the decay rate of the $H\to hh$ mode and the cross section of the double Higgs boson production via the gluon fusion 
process $gg \to hh$ at the LHC. 
We find that the one-loop correction to the $hhh$ coupling significantly change the prediction at the tree level to be
${\cal O}(100)\%$ level under the constraint from perturbative unitarity, triviality, vacuum stability and conditions to avoid wrong vacua.  
Furthermore, the cross section of $gg \to hh$ can be more than 20 times larger than the SM prediction 
due to the resonance effect of $H$. 

This paper is organized as follows. 
In Sec.~II, we define the Lagrangian of the HSM. 
In Sec.~III, we discuss bounds on the parameter space from theoretical and experimental constraints. 
In Sec.~IV, the renormalization of parameters in the Higgs potential is described based on the on-shell scheme~\cite{HSM_KKY}. 
Numerical analyses for the one-loop corrected $hhh$ coupling, the decay rate of $H\to hh$ and the cross section of the double Higgs boson production process via $gg\to hh$
are given in Sec.~V. 
Conclusions are summarized in Sec.~VI. 
In Appendices, we present the analytic expressions for the scalar triple and quartic couplings (Appendix~A), the one-loop beta functions for 
dimensionless coupling constants (Appendix~B) and the One Particle Irreducible (1PI) diagram contributions to the $hhh$ and $Hhh$ vertices (Appendix~C). 

\section{The Higgs Singlet Model}\label{sec:lag}

We define the Lagrangian of the HSM based on the $SU(2)_L\times U(1)_Y$ gauge theory, 
of which Higgs sector is composed of an isospin complex doublet scalar field $\Phi$ and 
an isospin real singlet scalar field $S$. 

The most scalar potential is given as
\begin{align}
V(\Phi,S) =& +m_\Phi^2|\Phi|^2+\lambda |\Phi|^4  
+\mu_{\Phi S}^{}|\Phi|^2 S+ \lambda_{\Phi S} |\Phi|^2 S^2 
+t_S^{}S +m^2_SS^2+ \mu_SS^3+ \lambda_SS^4,\label{Eq:HSM_pot}
\end{align} 
where the doublet and singlet fields can be parameterized by 
\begin{align}
\Phi=
\left[\begin{array}{c}
G^+\\
\frac{1}{\sqrt{2}}\big(v+\phi+iG^0\big)
\end{array}\right],\quad
S=v_S+s, 
\end{align}
with $G^+$ and $G^0$ being the Nambu-Goldstone bosons which are absorbed into the longitudinal components of the $W^+$ and $Z$ bosons, respectively. 
The Vacuum Expectation Value (VEV) of the singlet field $v_S^{}$ does not contribute to the electroweak symmetry breaking, so that the Fermi constant $G_F$ 
is determined only by the doublet VEV just like the SM: $v = (\sqrt{2}G_F)^{-1/2}\simeq 246$ GeV. 
Moreover, we can show that the shift of the singlet VEV does not change physics~\cite{CDL} as it is proved in the following\footnote{This is also true at the one-loop level, because 
the counter term of the singlet VEV $\delta v_S^{}$ can also be taken to be zero by reparametrizing the 
counter terms in the shifted Higgs potential $\delta V$ which is described by the same form as Eq.~(\ref{Eq:HSM_pot}), but all the parameters are replaced by those counter terms.
}. 
If we take the shift $S \to S + v_S' $, then the potential is rewritten by
\begin{align}
V(\Phi, S) & = (m_\Phi^2 + \mu_{\Phi S}^{}\, v_S' +  \lambda_{\Phi S}v_S^{\prime 2} )|\Phi|^2+ \lambda|\Phi|^4\notag\\
&+(\mu_{\Phi S}^{}+2\lambda_{\Phi S} v_S') |\Phi|^2 S + \lambda_{\Phi S}|\Phi|^2 S^2
+(t_S+2m_S^2v_S' +3\mu_S^{}v_S^{\prime 2}+4\lambda_S v_S^{\prime 3})S 
\notag\\
&+ (m_S^2+3\mu_Sv_S' +6\lambda_S v_S^{\prime 2})S^2+ (\mu_S + 4\lambda_Sv_S' )S^3 +\lambda_S S^4. 
\end{align}
Therefore, the modification of the potential by the shift $v_S'$ is absorbed by taking the following reparameterization:
\begin{align}
&m_\Phi^2 \to m_\Phi^2 - (\mu_{\Phi S}^{}\, v_S' +  \lambda_{\Phi S}v_S^{\prime 2} ), \notag\\
&\mu_{\Phi S}^{} \to \mu_{\Phi S}^{}-2\lambda_{\Phi S} v_S', \notag\\
&t_S \to t_S -(2m_S^2v_S' +3\mu_S^{}v_S^{\prime 2}+4\lambda_S v_S^{\prime 3}), \notag\\
&m_S^2 \to m_S^2-(3\mu_Sv_S' +6\lambda_S v_S^{\prime 2}), \notag\\
&\mu_S \to \mu_S - 4\lambda_Sv_S' . \label{shift}
\end{align}
Using this shift invariance, we can take $v_S^{}=0$ without loss of generality, and we set $v_S^{}=0$ in the following discussion to simplify expressions. 

The tadpole terms for $h$ and $s$ are respectively given by  
\begin{align}
T_\Phi = -v\left(m^2_\Phi +\lambda v^2 \right),\quad T_S =  -t_S^{}-\frac{1}{2}\mu_{\Phi S}^{}v^2.  \label{tp}
\end{align}
From the tadpole condition at the tree level; i.e., $T_\Phi=T_S=0$, 
we can eliminate $m_\Phi^2$ and $t_S^{}$. 
Under this condition, the mass terms in the potential are calculated as 
\begin{align}
V_{\text{mass}} &= \frac{1}{2}(s,\,\phi)
\begin{pmatrix}
M_{11}^2 & M_{12}^2 \\
M_{12}^2 & M_{22}^2
\end{pmatrix}
\begin{pmatrix}
s \\
\phi 
\end{pmatrix}, \label{mixing}
\end{align}
where 
\begin{align}
M^2_{11}= 2m_S^2+ v^2\lambda_{\Phi S}  ,\quad
M^2_{22}=2\lambda v^2,\quad M^2_{12}=v\mu_{\Phi S}^{}. \label{mij}
\end{align}
The mass eigenstates of two scalar bosons are defined by introducing the mixing angle $\alpha$ as 
\begin{align}
\begin{pmatrix}
s \\
\phi
\end{pmatrix} = R(\alpha)
\begin{pmatrix}
H \\
h
\end{pmatrix}~~\text{with}~~R(\theta) = 
\begin{pmatrix}
\cos\theta & -\sin \theta \\
\sin\theta & \cos\theta
\end{pmatrix}. 
\end{align}
We identify the mass eigenstate $h$ as the discovered Higgs boson at the LHC with the mass 125 GeV. 
In this basis, the mass matrix is diagonalized as follows
\begin{align}
R(\alpha)^T
\begin{pmatrix}
M_{11}^2 & M_{12}^2 \\
M_{12}^2 & M_{22}^2
\end{pmatrix}
R(\alpha) = \begin{pmatrix}
m_H^2 & 0 \\
0 & m_h^2
\end{pmatrix},
  \label{mixing2}  
\end{align}
From Eq.~(\ref{mixing2}), the mass eigenvalues and the mixing angle $\alpha$ are expressed by 
\begin{align}
 &m_H^2=M_{11}^2c^2_\alpha +M_{22}^2s^2_\alpha +M_{12}^2s_{2\alpha} , \label{mbh}\\
 &m_h^2=M_{11}^2s^2_\alpha +M_{22}^2c^2_\alpha -M_{12}^2s_{2\alpha} , \label{mh}\\
 &\tan 2\alpha=\frac{2M_{12}^2}{M_{11}^2-M_{22}^2},   \label{tan2a}
 \end{align} 
where we introduced the shorthand notation for the trigonometric functions: $c_\theta^{} = \cos \theta $ and $s_\theta^{}= \sin \theta$. 
Using Eqs.~(\ref{mbh})-(\ref{tan2a}), the parameters $\lambda$, $m_S^2$ and $\mu_{\Phi S}^{}$ can be rewritten by 
\begin{align}
\lambda &= \frac{1}{2v^2}(m_h^2c^2_\alpha + m_H^2 s^2_\alpha), \\
m_S^2  &= \frac{1}{2}\left(m_h^2s_\alpha^2 +  m_H^2c_\alpha^2 -\lambda_{\Phi S}v^2\right), \\
\mu_{\Phi S}^{} &= \frac{1}{v}s_\alpha c_\alpha\left(m_H^2-m_h^2 \right).  \label{aaa}
\end{align}
From the above discussion, the 7 independent parameters in the potential
are expressed by 
\begin{align}
&m_h,~~m_H,~~\alpha,~~v,~~\lambda_S,~~\lambda_{\Phi S},~~\mu_{S}. 
\end{align}
Among them, $m_h\simeq 125$ GeV and $v\simeq 246$ GeV are known parameters by experiments. 

It is important to mention here that the mixing angle $\alpha$
can also be expressed from Eq.~(\ref{aaa}) as 
\begin{align}
s_{2\alpha} &= \frac{2v\mu_{\Phi S}^{}}{m_H^2-m_h^2}. 
\end{align}
From this expression, we see that 
the mixing angle can be approximately given by $\alpha \simeq v/m_H$ when we consider the case for $\mu_{\Phi S}^{} \simeq m_H^{}$ and $m_H^{} \gg m_h$. 
Therefore, in this case the value of $\alpha$ is suppressed only by $1/m_H^{}$, i.e., instead of $1/m_H^2$.  
This feature is not seen in THDMs because of a lack of gauge invariant scalar trilinear couplings such as $\mu_{\Phi S}^{}$ in the HSM, 
where a mixing angle between two CP-even Higgs bosons is typically suppressed by the 
squared inverse of the mass of extra Higgs bosons. 
For this reason, the decoupling behavior by taking a large value of the mass of the extra Higgs boson $H$
is more slowly seen in the HSM as compared to that in THDMs. 

The kinetic Lagrangian ${\cal L}_{\text{kin}}$ and the Yukawa Lagrangian ${\cal L}_Y$ are given by 
\begin{align}
{\cal L}_\text{kin} &= |D_\mu \Phi|^2 +\frac{1}{2}(\partial_\mu S)^2 ,  \label{kin} \\
{\cal L}_Y         &= -y_d\bar{Q}_L\Phi d_R -y_u\bar{Q}_L i\sigma_2\Phi^* u_R -y_e\bar{L}_L\Phi e_R +\text{h.c.}, \label{yuk}
\end{align}
where $D_\mu$ is the covariant derivative for $\Phi$ and $\sigma_2$ is the second Pauli matrix. 
The trilinear interaction terms among the Higgs boson and SM particles are then extracted as
\begin{align}
{\cal L}_{\text{3-int}} & = \left(\frac{h}{v}c_\alpha +\frac{H}{v}s_\alpha \right)(2m_W^2 W^+_\mu W^{-\mu} + m_Z^2Z_\mu Z^\mu - m_f\bar{f}f). 
\end{align}
Therefore, the scaling factor  
of the Higgs boson coupling with weak bosons 
$\kappa_V^{}\equiv g_{hVV}^{\text{HSM}}/g_{hVV}^{\text{SM}}$
and fermions $\kappa_f^{}\equiv g_{hff}^{\text{HSM}}/g_{hff}^{\text{SM}}$
are universally given at the tree level by 
\begin{align}
\kappa_f = \kappa_V = c_\alpha. 
\end{align}
From the Higgs potential given in Eq.~(\ref{Eq:HSM_pot}), the scaling factor for the triple Higgs boson coupling $hhh$ is also calculated at the tree level as 
\begin{align}
\kappa_h \equiv \frac{\lambda_{hhh}}{\lambda_{hhh}^{\text{SM}}} = c_\alpha^3 +\frac{2v^2}{m_h^2}c_\alpha s_\alpha^2 \lambda_{\Phi S} - \frac{2v\mu_S}{m_h^2}s_\alpha^3. \label{kht}
\end{align}
We see from Eq.~(\ref{kht}) that $\kappa_h$ is more sensitive to the mixing angle as compared to $\kappa_f$ and $\kappa_V^{}$. 
All the tree level expressions for the scalar trilinear $\lambda_{\phi_1\phi_2\phi_3}$ and quartic
$\lambda_{\phi_1\phi_2\phi_3\phi_4}$ couplings are presented in Appendix~A.

\section{Constraints in the model}

We discuss constraints on the parameter space from theoretical arguments, namely, from 
perturbative unitarity~\cite{unitarity}, triviality~\cite{HSM-RGE,HSM-RGE2}, vacuum stability~\cite{HSM-RGE,HSM-RGE2} and wrong vacuum conditions~\cite{HSM-EWBG3}. 
We explain how the parameter space can be restricted by taking into account each of four constraints in order. 

First, the bound from perturbative unitarity is obtained by requiring that eigenvalues of the S-wave amplitude matrix for the 2 body to 2 body elastic scattering processes
are smaller than a given critical value~\cite{lqt}. 
In our model, all the eigenvalues are calculated at high energies by~\cite{unitarity} 
\begin{align}
x_1^\pm &= \frac{1}{16\pi}\left[3\lambda +6\lambda_S \pm \sqrt{(6\lambda_S-3\lambda)^2+4\lambda_{\Phi S}^2}\right], \label{eigen1}\\
x_2 & = \frac{1}{8\pi}\lambda,   \label{eigen2}  \\
x_3 & = \frac{1}{8\pi}\lambda_{\Phi S}   \label{eigen3}. 
\end{align}
For each of eigenvalues, we impose 
\begin{align}
|x_i| \leq 1/2. 
\end{align}

\begin{figure}[t]
\begin{center}
\includegraphics[width=80mm]{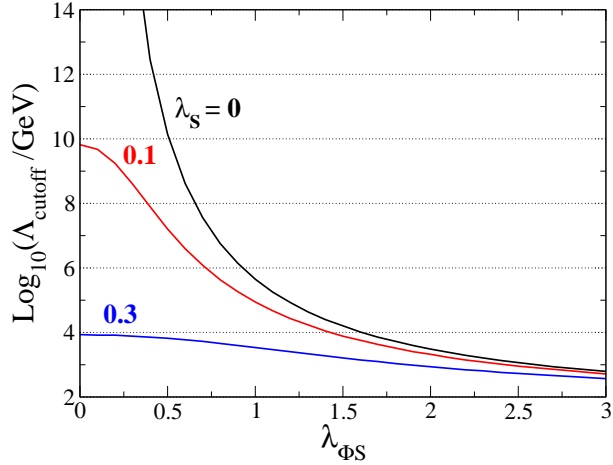}
\end{center}
\caption{Cutoff scale defined by Eq.~(\ref{cutoff1}) as a function of the initial value of $\lambda_{\Phi S}^{}$ at $m_Z^{}$
in the case of $\alpha=0$ and $m_t=173.21$ GeV. 
Each curve shows the different choice of $\lambda_S(m_Z^{})$. 
}
\label{beta-func}
\end{figure}

Second, the triviality bound is obtained by requiring that the Landau pole does not appear below a certain energy scale $\Lambda_{\text{cutoff}}$. 
Instead of using the scale where the Landau pole appears, we can define the triviality bound as follows
\begin{align}
\Big|\text{Max}[\lambda_i(\mu)] \Big| < 4\pi,~~ \text{for}~~^\forall\mu \leq \Lambda_{\text{cutoff}}, \label{cutoff1}
\end{align}
where $\lambda_i(\mu)$ are the scale dependent dimensionless coupling constants at a scale $\mu$. 
The scale dependence of $\lambda_i$ is calculated by solving the renormalization group equations (RGEs) 
for all the dimensionless coupling constants, and the full set of RGEs at the one-loop level are given in Appendix~B. 
Depending on  $\Lambda_{\text{cutoff}}$, we obtain the bound on $\lambda_i$ at the initial scale which is taken to be $m_Z^{}$. 
In Fig.~\ref{beta-func}, we show the cutoff scale as a function of $\lambda_{\Phi S}^{}(m_Z^{})$ for several fixed value of $\lambda_S^{}(m_Z^{})$ with $\alpha = 0$. 
We can see that the cutoff scale immediately becomes low when we take a non-zero value of $\lambda_S$, because of the large coefficient of the $\lambda_S^2$ term 
in the $\beta(\lambda_S)$ function given in Eq.~(\ref{lams}). 
We also see that a larger value of $\lambda_{\Phi S}$ makes the cutoff scale low, e.g., $\Lambda_{\text{cutoff}}\simeq 10^3~(3)$ TeV for $\lambda_{\Phi S}=1~(2)$ and $\lambda_{S}=0$. 
For $\alpha\neq 0$, the bound becomes stronger than that in the case with $\alpha=0$. 

Third, the constraint from the vacuum stability is imposed by requiring that 
the Higgs potential given in Eq.~(\ref{Eq:HSM_pot}) must be bounded from below in any direction with large scalar field values. 
This requirement can be expressed by 
\begin{align}
V^{(4)}(\Phi,S) \geq 0, \label{vs0}
\end{align}
where $V^{(4)}$ is the quartic term part of the potential. 
From Eq.~(\ref{vs0}), we obtain the following inequalities at a scale $\mu$~\cite{Fuyuto}:
\begin{align}
\lambda(\mu) \geq 0,\quad \lambda_S(\mu) \geq 0,
\quad 2\sqrt{\lambda(\mu) \lambda_S(\mu)} + \lambda_{\Phi S}(\mu) \geq 0,~~ \text{for}~~^\forall\mu \leq \Lambda_{\text{cutoff}}.  
\end{align}
If $\lambda_{\Phi S}(\mu)\geq 0 $, the last condition is trivial, while $\lambda_{\Phi S}(\mu) < 0 $, that is rewritten by 
\begin{align}
4\lambda(\mu) \lambda_S(\mu)\geq \lambda_{\Phi S}^2(\mu).  
\end{align}

Finally, we explain the bound from wrong vacuum conditions. 
In the HSM, because of the existence of the scalar trilinear couplings $\mu_S^{}$ and $\mu_{\Phi S}^{}$, 
non-trivial local extrema can appear in the Higgs potential.
Therefore, we have to check whether the true extremum at
$(\sqrt{2}\langle \Phi \rangle, \langle S \rangle) = (v_{\text{ew}}^{}, 0)$ with $v_{\text{ew}}\simeq 246$ GeV
corresponds to the minimum of the potential. 
According to Refs.~\cite{CDL,HSM-EWBG3}, the following five extrema appear 
 \begin{align}
 (\sqrt{2}\langle \Phi \rangle, \langle S \rangle) = (v_+,x_+), \,\,(v_-,x_-),\,\,
             (0,x_1^0),\,\,(0,x_2^0),\,\,(0,x_3^0),
 \end{align}
where
 \begin{align}
 x_{\pm} &\equiv \frac{3v_{\text{ew}}^{}(\mu_{\Phi S}^{}\lambda_{\Phi S}^{}-2\mu_S^{}\lambda)\pm2\sqrt{\Delta}}{4v_{\text{ew}}^{}(4\lambda\lambda_S^{}-\lambda_{\Phi S}^{2})}, \\
 v_{\pm}^2 &\equiv
 v_{\text{ew}}^2 -\frac{1}{\lambda}(\mu_{\Phi S}^{}x_\pm^{}+\lambda_{\Phi S}^{}x_\pm^2), \\
 x_1^0 &\equiv
 \frac{(6\mu_S^{}-\eta^{1/3})^2-96m_S^2 \lambda_S^{}}{24\lambda_S^{}\eta^{1/3}}
  +\frac{\mu_S^{}}{4\lambda_S^{}}, \\
 x_2^0 &\equiv
 \frac{(6\mu_S^{}-e^{2i\pi/3}\eta^{1/3})^2-96m_S^2 \lambda_S^{}}{24\lambda_S^{}e^{2i\pi/3}\eta^{1/3}}
  +\frac{\mu_S^{}}{4\lambda_S^{}}, \\
 x_1^0 &\equiv
 \frac{(6\mu_S^{}-e^{4i\pi/3}\eta^{1/3})^2-96m_S^2 \lambda_S^{}}{24\lambda_S^{}e^{4i\pi/3}\eta^{1/3}}
  +\frac{\mu_S^{}}{4\lambda_S^{}}, 
 \end{align}
with
 \begin{align}
 \Delta &\equiv
 \frac{9v_{\text{ew}}^2}{4}(2\mu_S^{}\lambda - \mu_{\Phi S}^{}\lambda_{\Phi S}^{})^2
 - 2 m_h^2 m_H^2 (4\lambda \lambda_S^{}- \lambda_{\Phi S}^2), \\
 \eta &\equiv
 12\left[  -18\mu_S^{}(\mu_S^2-4\lambda_S m_S^2) 
    + 72 \lambda_{S}^2 \mu_{\Phi S}^{}v_{\text{ew}}^2
    + \lambda_S^{}\sqrt{3\Delta_0^{}}\right], \\
 \Delta_0^{} &\equiv
 - 32\left[2m_S^4(9\mu_S^2 - 32\lambda_S m_S^2 )
 +27\mu_{\Phi S}^{} \mu_S^{} v_{\text{ew}}^2 (\mu_S^2 - 4m_S^2 \lambda_S^{})
 -54\lambda_S^2 \mu_{\Phi S}^2 v_{\text{ew}}^2\right].
 \end{align}
Now, the condition to avoid the wrong vacuum can be expressed by 
 \begin{align}
 V_{\text{nor}}(v_\pm,x_\pm) > 0, \,\,\,\,\,
 V_{\text{nor}}(0,x_{1,2,3}) > 0,
 \end{align}
where $V_{\text{nor}}$ is the normalized Higgs potential satisfying $V_{\text{nor}}(v_{\text{ew}}^{},0)=0$:
 \begin{align}
 V_{\text{nor}}(\phi,s) \equiv \frac{\lambda}{4}(\phi^2 - v_{\text{ew}}^2)^2
                + \left(\frac{\mu_{\Phi S}^{}}{2}s+ \frac{\lambda_{\Phi S}^{}}{2}s^2\right)(\phi^2 - v_{\text{ew}}^2)  
                + m_S^2  s^2
                + \mu_S^{} s^3 + \lambda_{S}^{}s^4.
 \end{align}

Before closing this section, we briefly comment on constraints from experimental data. 
In Ref.~\cite{mw_hsm,Robens,Robens2}, constraints from electroweak precision observables and Higgs boson search data at the LHC have been 
studied in the HSM with a spontaneously broken discrete $Z_2$ symmetry. 
It has been clarified that the constraint from the measurement of the W boson mass gives the strongest upper bound on $|s_\alpha|$ in 
the most of the parameter space. 
This bound becomes stronger when $m_H$ increases, e.g., $|s_\alpha| \lesssim 0.3~(0.2)$ for $m_H^{}=300~(800)$ GeV. 
We note that this bound can be applied to our model, because it only depends on $m_H^{}$ and $s_\alpha$. 

\section{Renormalization} \label{renormalization}

In this section, we calculate the renormalized scalar trilinear vertices $\hat{\Gamma}_{hhh}$ and $\hat{\Gamma}_{Hhh}$ at the one-loop level based on the 
on-shell scheme, where for some parameters we apply to the minimal subtraction scheme.  
The renormalized $hVV$ $(V=W,Z)$ and $hf\bar{f}$ vertices have already been calculated in Ref.~\cite{HSM_KKY}, so that we focus on the renormalization of the parameters in the 
Higgs potential. 
We first shift relevant parameters into the renormalized one and the counter term. 
We then give a set of renormalization conditions to determine these counter terms. 
In this paper, the calculations are performed in the 't~Hooft--Feynman gauge\footnote{It 
has been pointed out in Ref.~\cite{Robens_reno} that there remain gauge dependences in the mixing parameter $\alpha$ determined by the on-shell scheme. 
In Refs.~\cite{Robens_reno, Santos,THDM3,Pilaftsis}, a renormalization scheme to remove such a gauge dependence has been proposed. 
Although in our paper we apply to the usual on-shell renormalization scheme even if there remains the gauge dependence, 
the ratio of numerical values of physical observables such as the decay rate of the Higgs bosons 
calculated in the on-shell scheme to the improved scheme without the gauge dependence
has been known to be smaller than ${\cal O}(1)\%$~\cite{Robens_reno}. 
This means, for example, that if one-loop corrections to a quantity are given to be 1\%, the impact on the gauge dependence is less than ${\cal O}(0.01)\%$. }. 

\subsection{Shift of parameters}

The following eight bare parameters in the potential are shifted as
\begin{align}
T_{\Phi}&\to \delta T_{\Phi},\quad T_{S}\to \delta T_{S}, \notag\\
m_{h}^2 &\to m_{h}^2+\delta m_{h}^2,\quad
m_{H}^2 \to m_{H}^2+\delta m_{H}^2, \quad 
\alpha\to \alpha+\delta\alpha, \notag\\ 
\lambda_S^{} &\to \delta \lambda_S^{}, \quad \lambda_{\Phi S}^{} \to \delta \lambda_{\Phi S}^{},\quad \mu_{S}^{}\to  \delta\mu_S^{} .  \label{counter1}
\end{align}
In addition, the wave function renormalization for the scalar fields is given by the following way:
\begin{align}
\begin{pmatrix}
H \\
h
\end{pmatrix}
\to
\begin{pmatrix}
1+\frac{1}{2}\delta Z_H & \delta C_{Hh}+\delta\alpha\\
\delta C_{hH}-\delta\alpha  & 1+\frac{1}{2}\delta Z_h
\end{pmatrix}
\left(\begin{array}{c}
H\\
h
\end{array}\right),   \label{counter2}
\end{align}

Using the above counter terms, we can construct the renormalized scalar boson one- and two-point functions. 
In the following, we express 
contributions from 1PI diagrams as $\Gamma_\varphi^{\rm{1PI}}$ for the one-point scalar function and 
$\Pi_{\varphi\varphi'}^{\rm{1PI}}$ for the two-point scalar function. 
The renormalized one-point function for $h$ and $H$ are given by
\begin{align}
\hat{T}_h = \delta T_h + \Gamma_h^{\rm{1PI}},\quad 
\hat{T}_H = \delta T_H + \Gamma_H^{\rm{1PI}}, 
\end{align}
where
\begin{align}
\left(\begin{array}{c}
\delta T_S \\
\delta T_\Phi
\end{array}\right)=
R(\alpha)\left(\begin{array}{c}
\delta T_H\\
\delta T_h
\end{array}\right).
\end{align}
The renormalized two-point functions are expressed as
\begin{align}
\hat{\Pi}_{hh}(p^2)&=\Pi_{hh}^{\text{1PI}}(p^2)+\frac{c_\alpha^2\delta T_\Phi}{v}+\left[(p^2-m_h^2)\delta Z_h-\delta m_h^2\right],\\
\hat{\Pi}_{HH}(p^2)&=\Pi_{HH}^{\text{1PI}}(p^2)+\frac{s_\alpha^2\delta T_\Phi}{v}+\left[(p^2-m_H^2)\delta Z_H-\delta m_H^2\right],\\
\hat{\Pi}_{Hh}(p^2)&=\Pi_{Hh}^{\text{1PI}}(p^2)+ \frac{s_\alpha c_\alpha\delta T_\Phi}{v}
+p^2(\delta C_{hH}+\delta C_{Hh})+m_h^2(\delta \alpha -\delta C_{hH})-m_H^2(\delta \alpha +\delta C_{Hh}). 
\end{align}

\subsection{Renormalization conditions in the Higgs potential}

In the previous subsection, we prepared totally 12 counter terms from Eqs.~(\ref{counter1}) and (\ref{counter2}). 
We thus need 12 renormalization conditions to determine them. 
First, we impose two tadpole conditions at the one-loop level, i.e.,
\begin{align}
\hat{T}_h=\hat{T}_H=0. 
\end{align}
We then obtain 
\begin{align}
\delta T_h = -\Gamma_h^{\rm{1PI}},\quad  \delta T_H = -\Gamma_H^{\rm{1PI}}. 
\end{align}
Second, four on-shell conditions for the two-point functions: 
\begin{align}
&\hat{\Pi}_{hh}(m_h^2)=\hat{\Pi}_{HH}(m_H^2)=0,  \quad
\frac{d}{dp^2}\hat{\Pi}_{hh}(p^2)\big|_{p^2=m_h^2}=\frac{d}{dp^2}\hat{\Pi}_{HH}(p^2)\big|_{p^2=m_H^2}=0,
\label{rc2_s}
\end{align}
which determine the following four counter terms 
\begin{align}
\delta m_h^2&=\Pi_{hh}^{\text{1PI}}(m_h^2)+\frac{c_\alpha^2\delta T_\Phi}{v}, \quad
\delta m_H^2= \Pi_{HH}^{\text{1PI}}(m_H^2)+\frac{s_\alpha^2\delta T_\Phi}{v}, 
\end{align}
and 
\begin{align}
\delta Z_{h} &=-\frac{d}{d p^2}\Pi_{hh}^{\text{1PI}}(p^2)\Big|_{p^2=m_h^2}, \quad
\delta Z_{H} =-\frac{d}{d p^2}\Pi_{HH}^{\text{1PI}}(p^2)\Big|_{p^2=m_H^2}. \label{Zphi}
\end{align}
Three counter terms $\delta\alpha$, $\delta C_{hH}$ and $\delta C_{Hh}$ are determined by imposing the following three conditions
\begin{align}
&\hat{\Pi}_{Hh}(m_h^2)=\hat{\Pi}_{Hh}(m_H^2)=0, \quad \delta C_{hH}=\delta C_{Hh}\equiv\delta C_h, 
\end{align}
by which we obtain
\begin{align}
&\delta\alpha = \frac{1}{2(m_H^2-m_h^2)}\left[\Pi_{Hh}^{\text{1PI}}(m_h^2)+\Pi_{Hh}^{\text{1PI}}(m_H^2)+\frac{s_{2\alpha}\delta T_\Phi}{v}\right],\\
&\delta C_h = \frac{1}{2(m_H^2-m_h^2)}\left[\Pi_{Hh}^{\text{1PI}}(m_h^2)-\Pi_{Hh}^{\text{1PI}}(m_H^2)\right]. 
\end{align}

From the above discussion, we determine 9 counter terms, but there remain 3 undetermined ones: $\delta \lambda_{\Phi S}^{}$, 
$\delta \mu_S^{}$ and $\delta \lambda_{S}^{}$. 
Among the 3 counter terms, 
$\delta \lambda_S$ does not enter the following discussion, which appears in the renormalization of the scalar quartic vertices. 
For the remaining two counter terms $\delta \lambda_{\Phi S}^{}$ and $\delta \mu_{S}^{}$, we apply the minimal subtraction scheme in which they are 
determined so as to remove the ultra-violet (UV) divergent part of the one-loop correction to the $hhh$ and $Hhh$ vertices. 
We will further discuss the determination of these counter terms in the next subsection.

\subsection{Renormalized vertices}\label{reno_couplings}

The renormalized $hhh$ and $Hhh$ vertices are expressed as 
\begin{align}
\hat{\Gamma}_{hhh}(p_1^2,p_2^2,q^2)&=3!\lambda_{hhh}+\delta \Gamma_{hhh}+\Gamma_{hhh}^{\text{1PI}}(p_1^2,p_2^2,q^2), \\ 
\hat{\Gamma}_{Hhh}(p_1^2,p_2^2,q^2)&=2!\lambda_{Hhh}+\delta \Gamma_{Hhh}+\Gamma_{Hhh}^{\text{1PI}}(p_1^2,p_2^2,q^2),
\end{align}
where $\delta \Gamma_{{\cal H}hh}$ and $\Gamma^{\text{1PI}}_{{\cal H}hh}$ (${\cal H}=h$ or $H$)
are the contributions from the counter terms and the 1PI diagrams for the ${\cal H}hh$ vertices, respectively. 
The scalar three point couplings $\lambda_{hhh}$ and $\lambda_{Hhh}$ are given in Appendix~A. 
%
The counter-term contributions are expressed by 
\begin{align}
\frac{1}{3!}\delta\Gamma_{hhh}&=\delta\lambda_{hhh}+\frac{3}{2}\lambda_{hhh}\delta Z_h+\lambda_{Hhh} (\delta\alpha + \delta C_h) \notag\\
&= \frac{3}{2}\lambda_{hhh}\delta Z_h + \lambda_{Hhh}\delta C_h
 -\frac{c_\alpha^3}{2v}\delta m_h^2 \notag\\
& + \frac{c_\alpha}{2v^2}\left(m_h^2c_\alpha^2 - 2s_\alpha^2 v^2\lambda_{\Phi S} \right)\delta v^{}
+\frac{m_h^2-m_H^2}{2v}c_\alpha^2 s_\alpha \delta \alpha +\delta M,   \label{delhhh} \\ 
\frac{1}{2!}\delta\Gamma_{Hhh}&=\delta\lambda_{Hhh}
+3\lambda_{hhh}(\delta C_{h}-\delta \alpha) 
+2\lambda_{HHh} (\delta\alpha + \delta C_h) 
+ \lambda_{Hhh} \left(\delta Z_h+ \frac{1}{2}\delta Z_H  \right) \notag\\
&=+ \lambda_{Hhh} \left(\delta Z_h + \frac{1}{2}\delta Z_H  \right)+(3\lambda_{hhh} + 2\lambda_{HHh})\delta C_h  \notag\\
&-\frac{c_\alpha^2 s_\alpha }{v} \left(\delta m_h^2 + \frac{1}{2}\delta m_H^2\right)
 +\frac{m_H^2-m_h^2}{8v}(c_{3\alpha}-5c_\alpha) \delta \alpha \notag\\
& +\frac{s_\alpha}{2v^2}[(2m_h^2 + m_H^2)c_\alpha^2  +v^2\lambda_{\Phi S} (1 + 3c_{2\alpha})] \delta v
 + \delta M' , 
 \label{GamHhh}
\end{align}
where $\delta M$ and $\delta M'$ are the undetermined counter term from the on-shell conditions which are expressed by 
\begin{align}
\delta M &=  \delta\mu_S^{}s_\alpha^3  - vc_\alpha s_\alpha^2 \delta \lambda_{\Phi S},\quad 
\delta M' =  -3c_\alpha s_\alpha^2 \delta \mu_S -\frac{v}{4}(s_\alpha -3s_{3\alpha})\delta \lambda_{\Phi S}. 
\end{align}
Applying the minimal subtraction scheme which is discussed in the previous subsection 
to $\delta M$ and $\delta M'$, we obtain 
\begin{align}
\delta M  &= -\frac{s_\alpha^2}{16\pi^2}\Big[\sum_f\frac{2N_c^fm_f^2}{v}\lambda_{\Phi S}c_\alpha -\frac{2c_\alpha^3}{v^3}(2m_W^4+m_Z^4)
-\frac{3}{v}\lambda_{\Phi S}c_\alpha(2m_W^2+m_Z^2)\notag\\
&+\frac{m_h^2}{4v}\lambda_{\Phi S}(11c_\alpha+c_{3\alpha})
+\frac{m_H^2}{v}\lambda_{\Phi S}c_\alpha s_\alpha^2
+4v\lambda_{\Phi S}(3\lambda_S + \lambda_{\Phi S}) c_\alpha
-36\mu_S\lambda_S s_\alpha
\Big]\Delta_{\text{div}},  \\
\delta M' &= \frac{s_\alpha}{16\pi^2}\Big[\sum_f\frac{N_c^fm_f^2}{v}\lambda_{\Phi S}(1+3c_{2\alpha}) 
-\frac{2m_W^4+m_Z^4}{v^3}c_\alpha^2(c_{2\alpha}-3)\notag\\
&-\frac{3(2m_W^2+m_Z^2)}{2v}\lambda_{\Phi S}(1+3c_{2\alpha})
+\frac{3m_h^2}{2v}\lambda_{\Phi S}c_\alpha^2(3 + c_{2\alpha})
-\frac{3m_H^2}{v}\lambda_{\Phi S} s_\alpha^4\notag\\
&+2v\lambda_{\Phi S}(3\lambda_S + \lambda_{\Phi S})(1+3c_{2\alpha})
-108\mu_S\lambda_S c_\alpha s_\alpha
\Big]\Delta_{\text{div}}, 
\end{align}
where $\Delta_{\text{div}}$ expresses the UV divergent part of the loop integral, and $N_c^f$ is the color factor; i.e., $N_c^f=3~(1)$ for $f$ being quarks (leptons). 

We note that the counter term of the VEV $\delta v$ is determined by using the gauge boson two-point functions which have been given in Ref.~\cite{HSM_KKY}. 

\section{Numerical results}

In this section, we perform the numerical analysis of some observables, i.e., 
the deviation in the $hhh$ coupling at the one-loop level from the SM prediction (Sec.~VA), 
the total width and the decay branching ratio of $H$ (Sec.~VB) 
and the double Higgs boson production cross section via the gluon fusion $gg\to hh$ at the LHC (Sec.~VC)
by using the one-loop renormalized $hhh$ and $Hhh$ vertices. 
In order to constrain the parameter space, we take into account the perturbative unitarity, 
triviality, vacuum stability and wrong vacuum conditions as we have explained in Sec.~III. 
The triviality and vacuum stability bound depend on the cutoff scale $\Lambda_{\text{cutoff}}$ of the model which is taken to be 
3 TeV or 10 TeV in the following analysis. 
In some plots shown in the following subsections, we also consider the constraint from the electroweak precision test for 
the $W$ boson mass in Ref.~\cite{Robens} which 
gives the upper limit on $|s_\alpha|$ for a given value of $m_H^{}$. 

For the numerical analysis, we have the following five free parameters 
\begin{align}
m_H,~s_\alpha,~\lambda_{\Phi S}^{},~\lambda_S^{},~\mu_S^{}. 
\end{align}
As we have seen in Fig.~\ref{beta-func}, a non-zero value of $\lambda_S$ significantly reduces the cutoff scale because of the RGE evolution of $\lambda_S^{}$. 
We thus simply take $\lambda_S^{}=0$ throughout this section to have the cutoff scale to be above the multi-TeV scale. 

We use the following SM input parameters~\cite{PDG}
\begin{align}
&\alpha_{\text{em}} =(137.035999074)^{-1},~
m_Z^{}=91.1876~\text{GeV},~
G_F=1.1663787\times 10^{-5}~\text{GeV}^{-2},   \notag\\
&\Delta\alpha_{\text{em}}=0.06635,~\alpha_s=0.1185~\text{GeV}, \notag\\ 
&m_t = 173.21~\text{GeV},~
m_b=4.66~\text{GeV},~
m_c=1.275~\text{GeV},~
m_\tau=1.77684~\text{GeV},~
m_h=125~\text{GeV}, 
\end{align}
where $\Delta\alpha_{\text{em}}$ is the shift of the fine structure constant which appears in the calculation of 
the photon self-energy by $d\Pi_{\gamma\gamma}^{\text{1PI}}(0)/dp^2=\Pi_{\gamma\gamma}^{\text{1PI}}(m_Z^2)/m_Z^2 +\Delta \alpha_{\text{em}}$ (see: e.g., \cite{Hollik,THDM3}). 

\subsection{One-loop corrected $hhh$ coupling}

\begin{figure}[t]
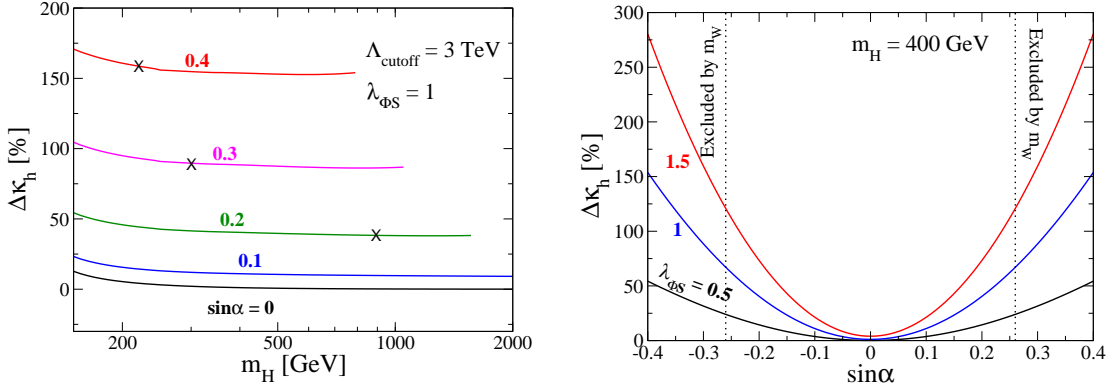

\begin{center}
\includegraphics[width=70mm]{hhh-mH_lamps1.eps}\hspace{5mm}
\includegraphics[width=70mm]{hhh-alpha_mH400.eps}
\caption{
Deviation in the $hhh$ coupling $\Delta\kappa_h$ as a function of $m_H$ (left) and $\sin\alpha$ (right) in the case of 
$\mu_S=0$ and $\Lambda_{\text{cutoff}}=3$ TeV. 
In the left panel, we take $\lambda_{\Phi S} = 1$ and $s_\alpha=0$, 0.1, 0,2, 0.3 and 0.4, while in the right panel we take $m_H=400$ GeV and $\lambda_{\Phi S}=0.5$, 1 and 1.5.  
The truncated point of the curve with $s_\alpha=0.2$, 0.3 and 0.4 in the left panel indicates the upper limit on $m_H^{}$
from the theoretical constraints. 
Besides, the point X shows the upper limit on $m_H^{}$ from $m_W$~\cite{Robens}. 
In the right panel, the vertical lines show the upper limit on $|\sin\alpha|$ from $m_W$. 
}
\label{1}
\end{center}
\end{figure}

The scaling factor of the $hhh$ coupling $\kappa_h$ is defined in Eq.~(\ref{kht}) at the tree level. 
Now, we grade up this quantity at the one-loop level as follows: 
\begin{align}
\kappa_h \equiv \frac{\hat{\Gamma}_{hhh}(m_h^2,m_h^2,4m_h^2)_{\text{HSM}}}{\hat{\Gamma}_{hhh}(m_h^2,m_h^2,4m_h^2)_{\text{SM}}}. 
\end{align}
Using this, the deviation in the $hhh$ coupling is expressed by $\Delta \kappa_h = \kappa_h -1$. 

First of all, we show the simple plot of $\Delta \kappa_h$ in Fig.~\ref{1}. 
The left and right panel respectively shows the $m_H$ and $s_\alpha$ dependence of $\Delta\kappa_h$. 
By looking at the curve with $s_\alpha = 0$ in the left panel, we can see the decoupling behavior of the 
$H$ loop effect to the $hhh$ coupling, i.e., the prediction asymptotically approaches to the SM value $(\Delta\kappa_h=0)$ as $m_H^{}$ is getting large.  
On the other hand, if we take $s_\alpha\neq 0$, the upper limit on $m_H^{}$ appears because of the theoretical constraints and the bound from $m_W$, so that we cannot 
take the decoupling limit in this case. 
It is also seen that in the region $m_H\gtrsim 300$ GeV, the prediction of $\Delta\kappa_h$ does not change so much.  
When we look at the right panel, we can see that $\Delta\kappa_h$ monotonically increases as $|s_\alpha|$ becomes large. 
Because of the bound from $m_W^{}$, we can extract the maximal allowed value of $\Delta\kappa_h$ to be about 120\%, 70\% and 20\% for $\lambda_{\Phi S}=1.5$, 1.0 and 0.5, respectively. 

\begin{figure}[t]
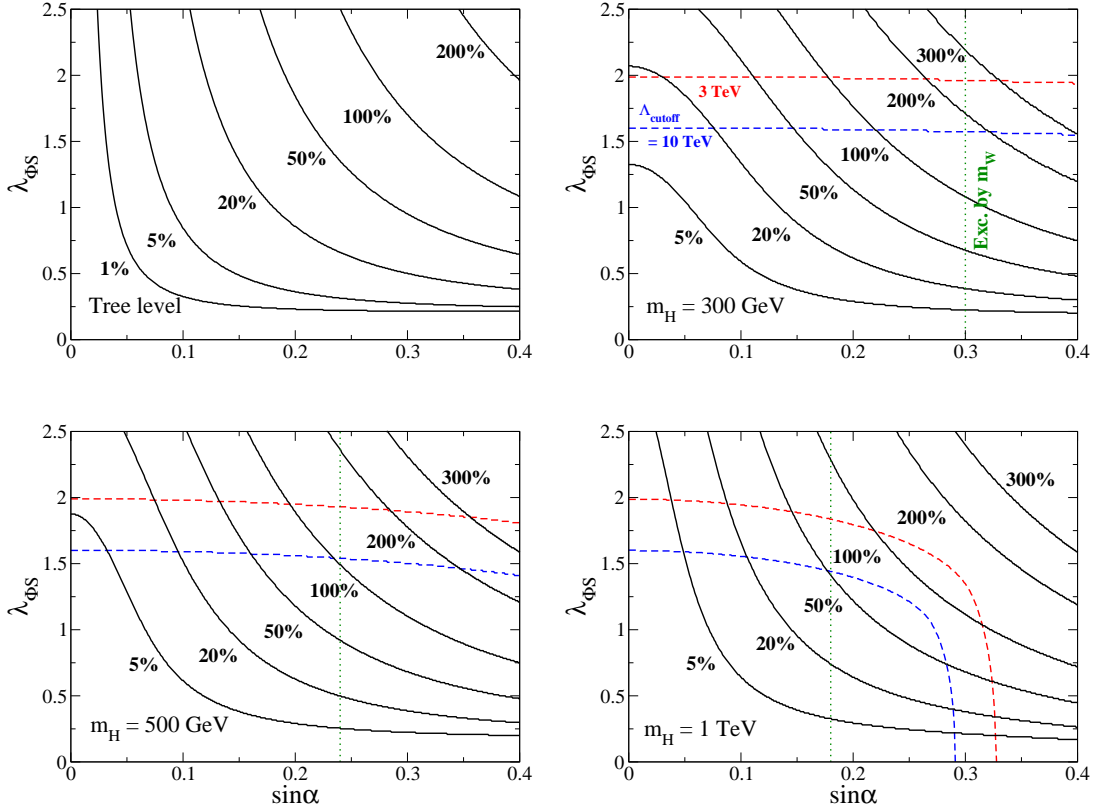

\begin{center}
\includegraphics[width=70mm]{contour-hhh_tree.eps}\hspace{3mm}
\includegraphics[width=70mm]{contour-hhh_mH300.eps}\\ \vspace{5mm}
\includegraphics[width=70mm]{contour-hhh_mH500.eps}\hspace{3mm}
\includegraphics[width=70mm]{contour-hhh_mH1000.eps}
\caption{
Contour plots for $\Delta\kappa_h$ on the $\sin\alpha$-$\lambda_{\Phi S}$ plane in the case of $\mu_S=0$. 
The upper-left (upper-right and lower) panel shows the tree level (one-loop corrected) result. 
In the upper-right, lower-left and lower-right panels, we take $m_H^{}=300$, 500 and 1000 GeV, respectively. 
The region above the blue (red) dashed curve is excluded by the theoretical constraints with $\Lambda_{\text{cutoff}}=10\,(3)$ TeV, while 
the right region from the vertical dotted line is excluded by $m_W$~\cite{Robens}. 
}
\label{2}
\end{center}
\end{figure}

In Fig.~\ref{2}, we show the contour plots for $\Delta\kappa_h$ on the $s_\alpha$-$\lambda_{\Phi S}$ plane. 
The upper-left panel shows the tree level result to see how the loop correction modifies the prediction, where 
$\Delta\kappa_h$ does not depend on $m_H^{}$ as it is shown in Eq.~(\ref{kht}). 
The upper-right and lower panels show the one-loop corrected results with $m_H^{}=300$ (upper-right), 500 (lower-left) and 1000 GeV (lower-right). 
By comparing the tree and one-loop corrected results, we see that the one-loop correction changes the tree level result to be ${\cal O}(100)\%$. 
The value of $\Delta \kappa_h$ becomes larger when $s_\alpha$ and/or $\lambda_{\Phi S}$ is getting large  at the both tree level and the one-loop level. 
The maximal allowed value of $\Delta\kappa_h$ can be extracted from the lower panels to be about 
250\%, 150\% and 75\% for $m_H^{}=300$, 500 and 1000 GeV, respectively, in the case of $\Lambda_{\text{cutoff}}=3$ TeV.  

We note that such a large correction to the $hhh$ coupling happens due to the non-decoupling effect of the $H$ loop when $m_H^{}$ mainly comes from 
the Higgs VEV\footnote{Although this intuitively seems to be a breaking of the perturbation theory, this does not follow the usual perturbative expansion. 
Namely, the magnitude of the one-loop $H$ loop contribution to the $hhh$ coupling depends on the $Hhh$ coupling which is independent of the $hhh$ coupling. 
Therefore, the amount of the correction does not simply follow the power expansion of the $hhh$ coupling with the loop factor. }.  
In the HSM, this non-decoupling effect appears in the case with $\lambda_{\Phi S} v^2 \gtrsim m_S^2$ or equivalently $\lambda_{\Phi S}={\cal O}(1)$ 
as we can see it in Eq.~(\ref{mij}). 
Similar non-decoupling effects in the $hhh$ coupling have also been found in the THDM as these have been pointed it out in Ref.~\cite{KOSY}. 
In fact, even in the SM
the non-decoupling effect in the $hhh$ can be seen on the top loop contribution as its mass purely comes from $v$, where
the magnitude of the correction is proportional to $m_t^4$, and it can be of order 10\% level. 
It goes without saying that the top loop effect is included in our calculation, but it does not change the value of $\Delta \kappa_h$ so much, 
because it is defined by the deviation from the SM prediction.

\begin{figure}[t]
\begin{center}
\includegraphics[width=70mm]{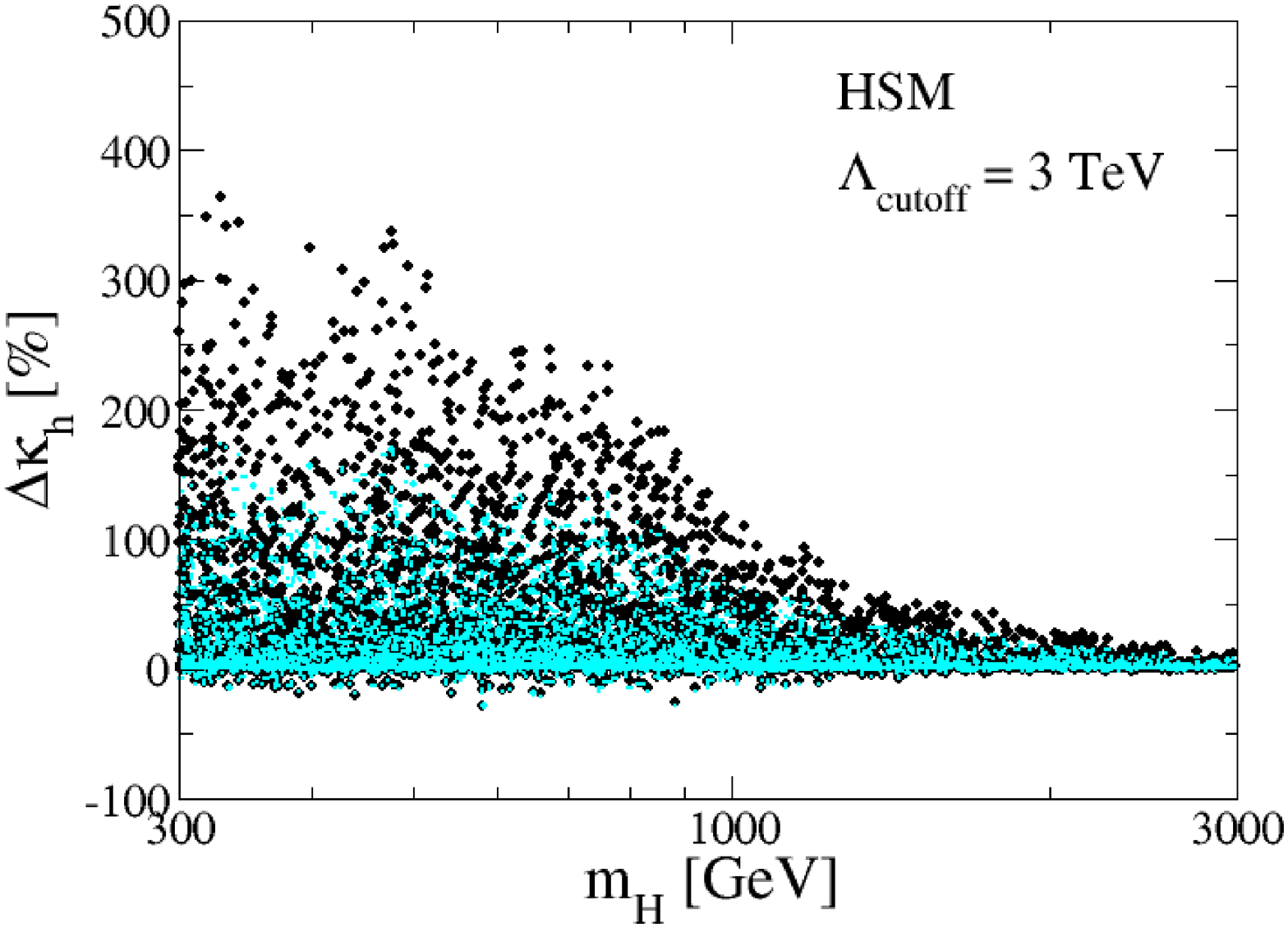} 
\includegraphics[width=70mm]{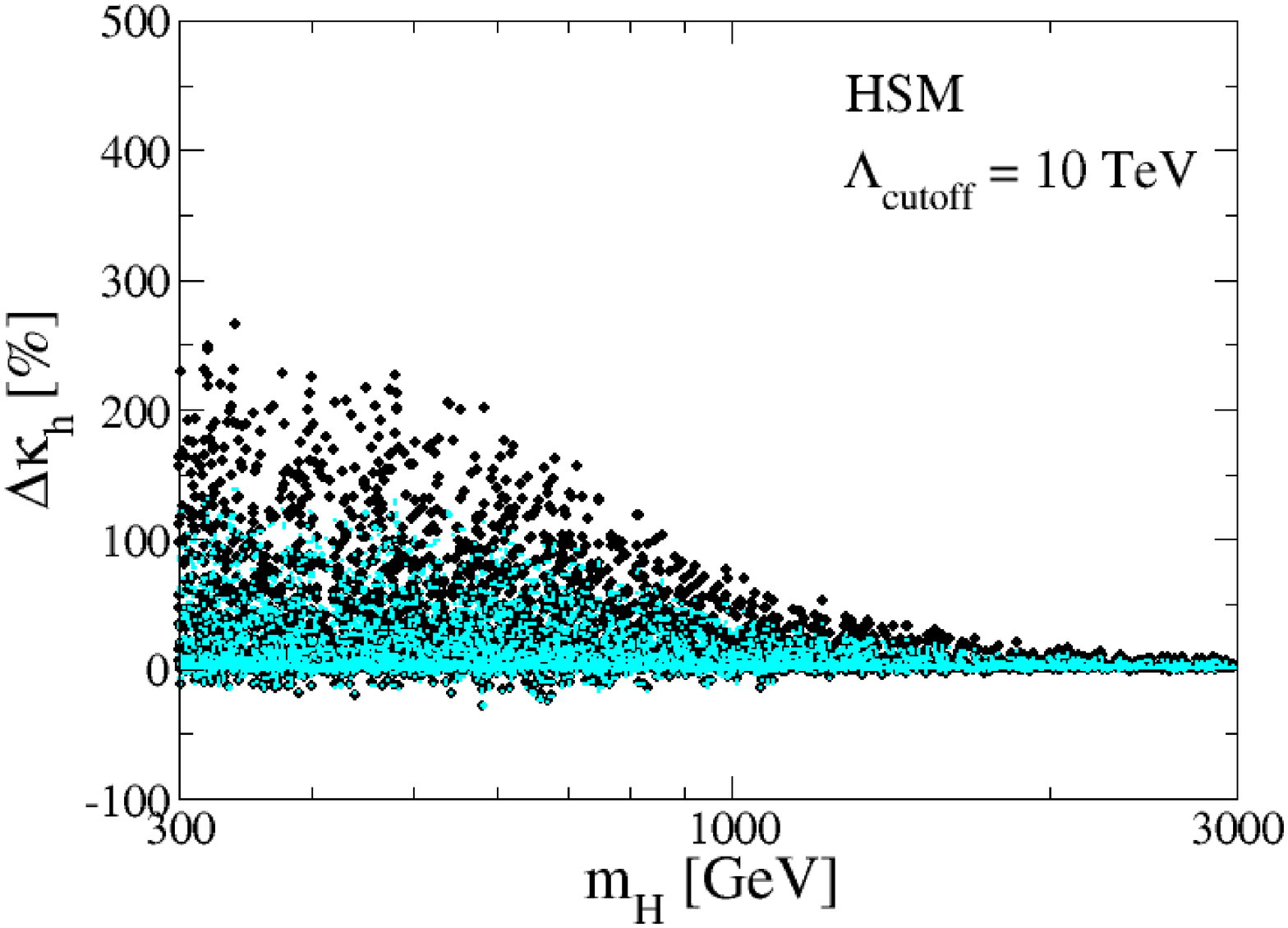} \\ \vspace{5mm}
\includegraphics[width=70mm]{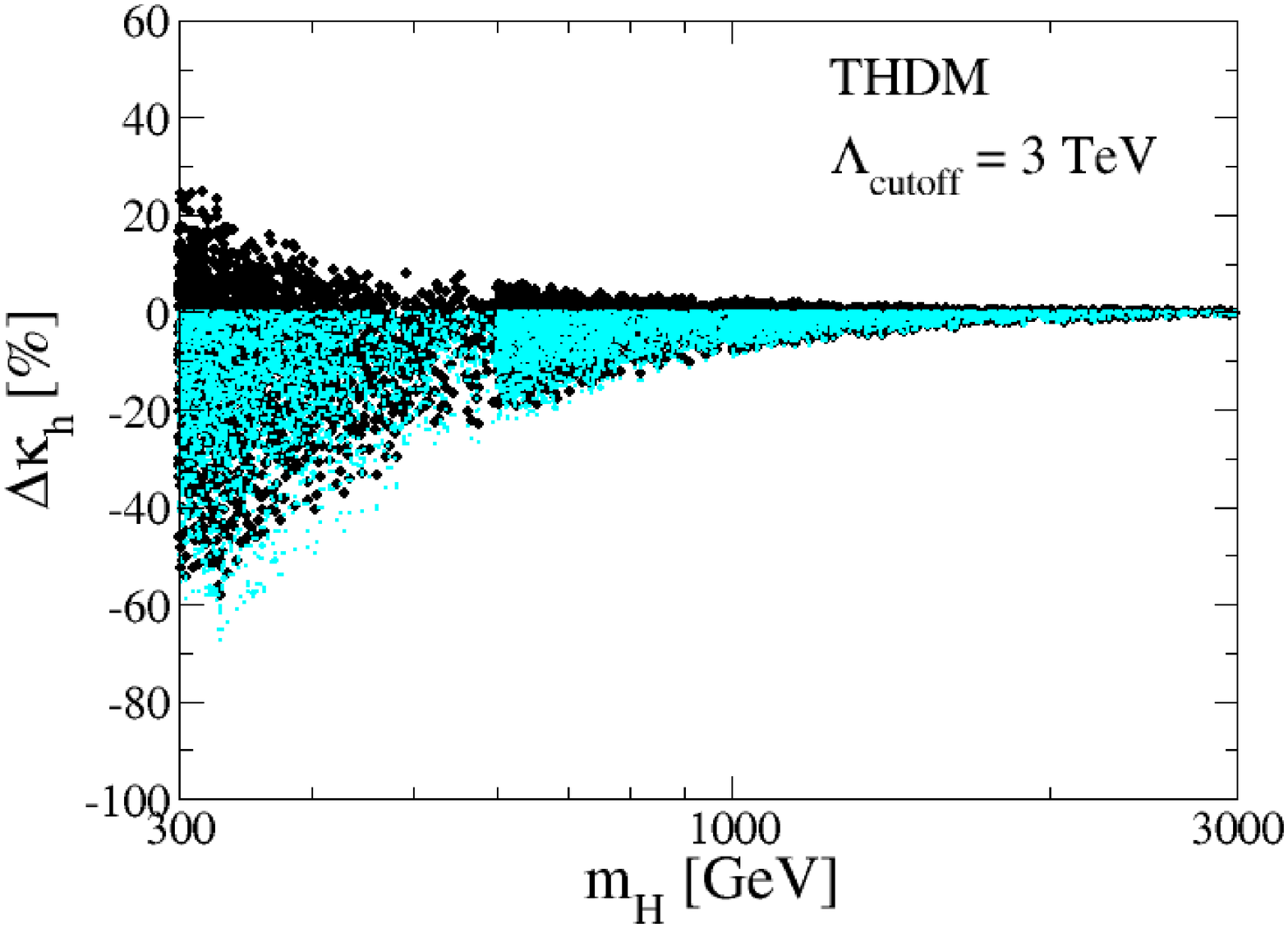} 
\includegraphics[width=70mm]{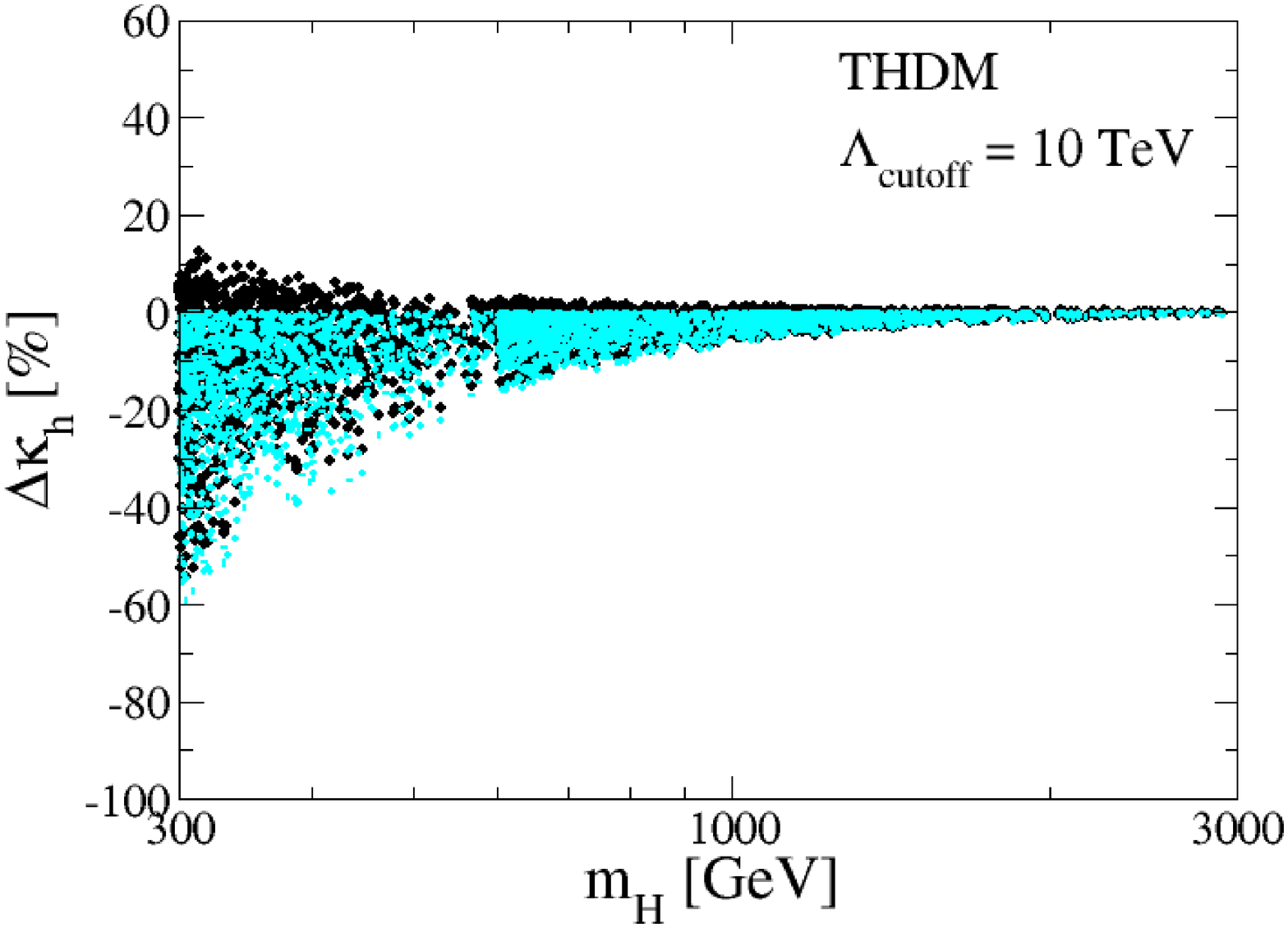}
\caption{
Scatter plot on the $m_H^{}$-$\Delta\kappa_h$ plane in the HSM (upper panels) and the Type-I THDM with $\tan\beta =1$ (lower panels). 
Each black (light blue) dot is the prediction allowed by theoretical constraints at the one-loop (tree) level. 
In the left and the right panels, we impose all the theoretical constraints assuming $\Lambda_{\text{cutoff}}=3$ and 10 TeV, respectively. 
}
\label{3}
\end{center}
\end{figure}

Finally, we scan the parameter space to see the possible allowed range of $\Delta\kappa_h$. 
In order to see the difference between the HSM and the THDM, we also calculate the one-loop corrected $hhh$ coupling based on the 
previous our work given in Ref.~\cite{THDM3}. 
We take the following scan range for the parameters
\begin{align}
&300 <m_H^{}< 3000~\text{GeV},~0<|s_\alpha|<0.4,~0<\lambda_{\Phi S}<2.5,~-100<\mu_{S}<100~\text{GeV},~\text{for~HSM}, \label{scan1}\\
&300 <m_\Phi^{}< 3000~\text{GeV},~0.92< s_{\beta-\alpha}<1,~0<\lambda_{\Phi\Phi h }/v<2.5,~\text{for~THDM},  \label{scan2}
\end{align}
where $m_\Phi^{}=m_H\,(=m_A^{}=m_{H^\pm})$, and 
$\lambda_{\Phi \Phi h}$ is defined by $v\lambda_{\Phi \Phi h}=  m_\Phi^2-M^2$. 
For the details of the definition of the parameters in the THDM, see, e.g., Ref.~\cite{KOSY}. 
We take into account the current bound on the Higgs boson couplings given at the LHC Run-I experiments. 
The combined results for the measurements of the scaling factors $\kappa_X^{}$ at the ATLAS and the CMS experiments have been provided in Ref.~\cite{combine} as follows
\begin{align}
\kappa_Z^{}    = 1.00^{+0.11}_{-0.10},\quad 
\kappa_W^{}    = 0.91^{+0.10}_{-0.12},\quad 
\kappa_\tau^{} = 0.90^{+0.14}_{-0.16},\quad 
\kappa_t^{}    = 0.87^{+0.15}_{-0.15}, 
\end{align}
where we pick up the positive allowed values of $\kappa_X^{}$. We require the predictions of the above scaling factors being inside the 2$\sigma$ level. 

In Fig.~\ref{3}, we show the scatter plot using the scanned parameter range given in Eqs.~(\ref{scan1}) and (\ref{scan2}) in the HSM (upper panels) 
and in the Type-I THDM with $\tan\beta=1$ (lower panels). 
In these plots, the constraint from $m_W$ is not imposed. 
The black (light blue) dot shows the allowed prediction at the one-loop (tree) level. 
By comparing the results in the HSM and in the THDM, we can find the big difference in the speed of the decoupling. 
Namely, the deviation in the $hhh$ coupling is more quickly shrunk as $m_H^{}$ increases in the THDM as compared to the HSM. 
This can be explained by the existence of the gauge invariant scalar trilinear coupling $\mu_{\Phi S}^{}$ in the HSM which makes the speed of the decoupling behavior
slow as we have discussed it in Sec.~II.   
For example, when we look at the region with $m_H^{}>1$ TeV, $\Delta\kappa_h$ can still be ${\cal O}(100)\%$ in the HSM, but it is a few percent level in the THDM. 
In addition, the value of $\Delta\kappa_h$ tends to be positive (negative) in the HSM (THDM). 
This difference can be explained by the difference of the structure of the sharing of VEVs and the mixing of CP-even Higgs fields at the tree level. 
In the HSM, the $hhh$ coupling can be decreased due to the non-zero field mixing between $h$ and $H$, but such reduction can be compensated by 
the additional contribution of the $\lambda_{\Phi S}$ parameter as it is seen in Eq.~(\ref{kht}). 
As a result, $\Delta \kappa_h$ at the tree level can typically be positive. 
On the other hand in the THDM, 
the $hhh$ coupling can be decreased not only by the field mixing but also by the sharing of VEVs such as $v_1^2+v_2^2=v^2$ ($v_{1,2}$ are VEVs of two Higgs doublets),
where the latter does not happen in the HSM. 
This additional reduction by the VEV mixing makes the $hhh$ coupling small as compared to the SM prediction. 
For reference, we give the tree level expression for $\kappa_h$ in the THDM:
\begin{align}
\kappa_h = s_{\beta-\alpha} -\frac{2(M^2 -m_h^2)}{m_h^2}s_{\beta-\alpha}c_{\beta-\alpha}^2 - \frac{M^2-m_h^2}{m_h^2}c_{\beta-\alpha}^3(\cot\beta -\tan\beta). 
\end{align}

From this result, when the second Higgs boson is discovered, and its mass is measured at future collider experiments, 
we can expect that the big difference in the value of the $hhh$ coupling appears between the HSM and the THDM. 

\subsection{Decay of $H$}

Here, we discuss the width of $H$ which is needed to calculate the cross section of the double Higgs boson production as it will be discussed in the next subsection. 
The total width of $H$ is calculated by 
\begin{align}
\Gamma_H = s_\alpha^2\Gamma_{h_{\text{SM}}}\big|_{m_{h_{\text{SM}}} \to m_H^{}} + H(m_H^{} - 2m_h)\times \Gamma(H\to hh), 
\end{align}
where $H(x) = 1~(0)~~\text{for}~~x\geq 0~(x<0)$. 
In the above formulae, the first term $\Gamma_{h_{\text{SM}}}\big|_{m_{h_{\text{SM}}} \to m_H^{}}$ represents the total width of 
the SM Higgs boson ${h_{\text{SM}}}$ but the mass $m_{h_{\text{SM}}}$ is replaced by $m_H^{}$. 
The second term corresponds to the partial width of $H \to hh$ decay mode which opens for $m_H \geq 2m_h = 250$ GeV. 
The analytic expression of the decay rate of $H \to hh$ is given by 
\begin{align}
\Gamma(H \to hh) = \frac{1}{32\pi m_H^{}}|\hat{\Gamma}_{Hhh}(m_h^2,m_h^2,m_H^2)|^2\sqrt{1-\frac{4m_h^2}{m_H^2}},~~\text{for}~~m_H^{} \geq 2m_h. 
\end{align}

\begin{figure}[t]
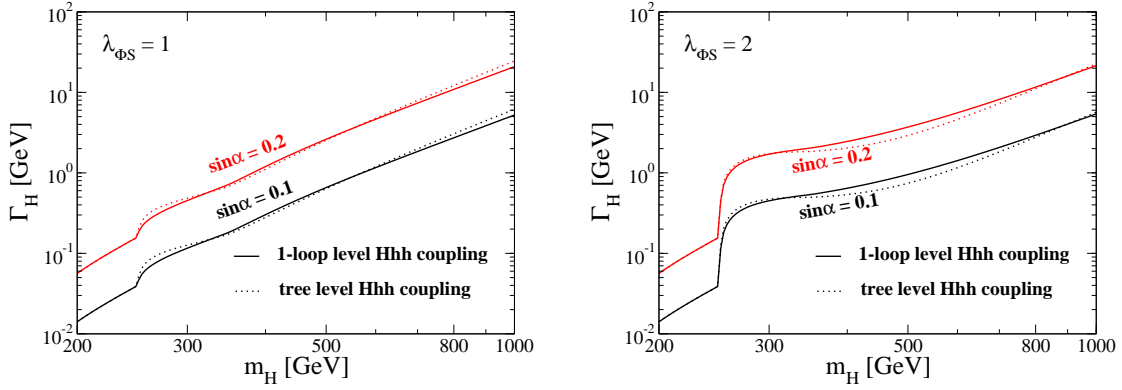

\begin{center}
\includegraphics[width=70mm]{width_H_lam1.eps} \hspace{5mm}
\includegraphics[width=70mm]{width_H_lam2.eps} \hspace{5mm} 
\caption{
Total width $\Gamma_H$ as a function of $m_H^{}$ in the case of $\mu_S=0$ and $\lambda_{\Phi S} = 1$ (left) and 2 (right). 
The solid and dotted curve denote the result using the one-loop corrected $Hhh$ vertex and the tree level one, respectively.
}
\label{width}
\end{center}
\end{figure}

In Fig.~\ref{width}, we show the $m_H^{}$ dependence of $\Gamma_H$. 
The solid and dotted curve show the result using the one-loop corrected $Hhh$ vertex ($\hat{\Gamma}_{Hhh}$) and the tree level vertex ($2\lambda_{Hhh}$), respectively. 
We can see the rapid growth of $\Gamma_H$ at around $m_H = 250$ GeV because of opening the channel $H \to hh$, where 
the amount of the growth with $\lambda_{\Phi S}=2$ is larger than that with $\lambda_{\Phi S}=1$. 
At large values of $m_H^{}$, the difference between the case for $\lambda_{\Phi S}=1$ and 2 becomes small. 
We also see that $\Gamma_H$ in the case of $s_\alpha=0.2$ is almost 4 times larger than that in the case of $s_\alpha=0.1$. 
The typical value of $\Gamma_H$ is found to be 10-100 MeV level when $m_H<2m_h$, while it becomes 1-10 GeV level when $m_H> 2m_h$.  

\begin{figure}[t]
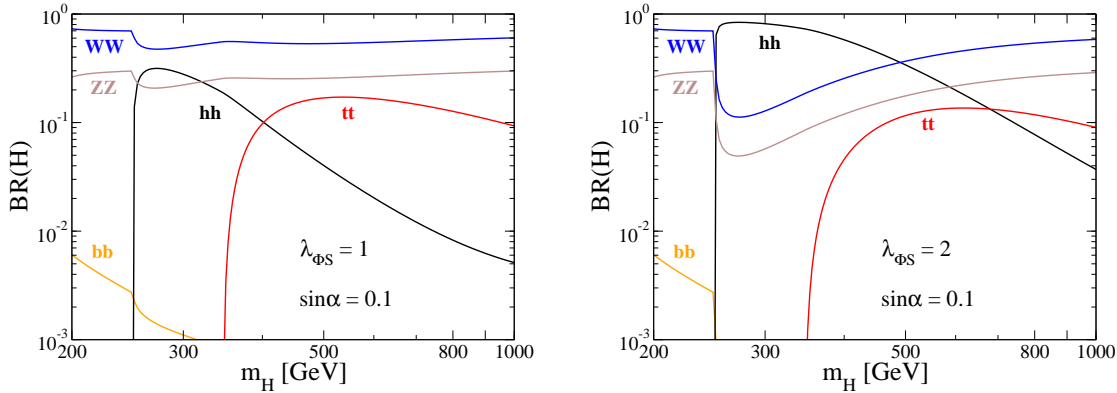

\begin{center}
\includegraphics[width=70mm]{br_H_lam1_sa01.eps} \hspace{5mm}
\includegraphics[width=70mm]{br_H_lam2_sa01.eps} \hspace{5mm} 
\caption{
Branching ratios BR$(H)$ as a function of $m_H^{}$ in the case of $\mu_S=0$, $\sin\alpha=0.1$ and $\lambda_{\Phi S} = 1$ (left) and 2 (right).
In this plot, we use the one-loop corrected $Hhh$ vertex. 
}
\label{br}
\end{center}

\end{figure}

In Fig.~\ref{br}, we show the $m_H^{}$ dependence of branching ratio of $H$ BR($H$) in the case of $s_\alpha = 0.1$. 
We note that the $s_\alpha$ dependence of the branching ratio is negligibly small. 
We can see that the di-Higgs boson channel $H\to hh$ with  $\lambda_{\Phi S}=2$ 
can be more important than that with $\lambda_{\Phi S}=1$, but the branching ratio of this channel becomes small as $m_H^{}$ increases. 

\subsection{Double Higgs boson production }

\begin{figure}[t]
\begin{center}
\includegraphics[width=150mm]{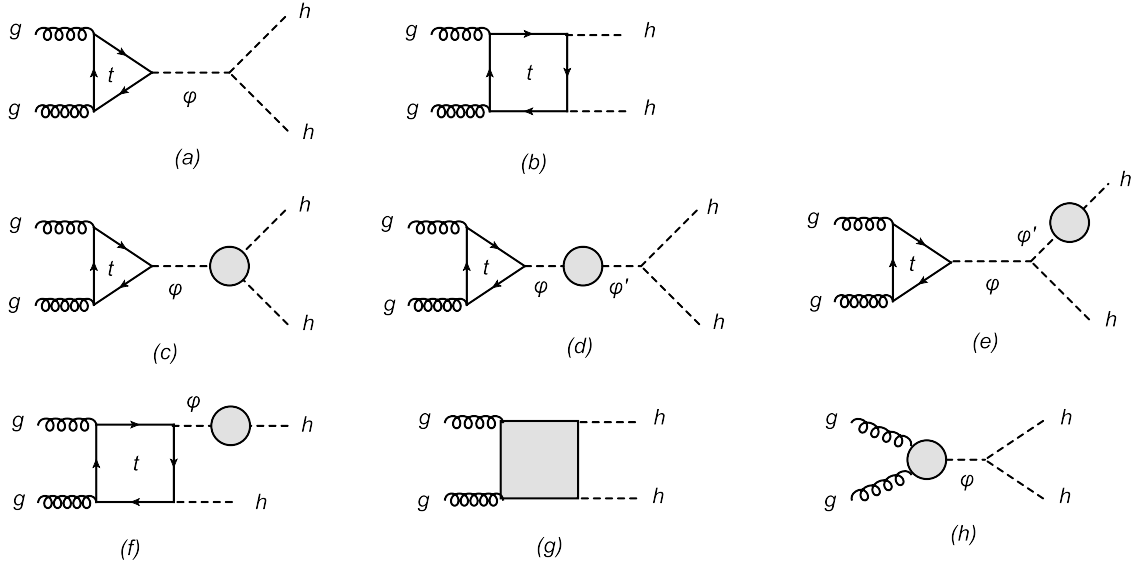}
\caption{
Feynman diagrams for the $gg \to hh$ process at the two-loop level with the order of $\alpha_s$. 
The gray blob in the diagram (c) and the diagrams (d), (e) and (f)
shows the insertion of 
the renormalized scalar three-point vertices $\hat{\Gamma}_{\varphi hh}^{\text{loop}}\equiv \delta \Gamma_{\varphi hh} + \Gamma_{\varphi hh}^{\text{1PI}}$ 
and the renormalized scalar two-point functions  $\hat{\Pi}_{\varphi \varphi'}$, respectively, where  
the symbols $\varphi$ and $\varphi'$ denote $h$ or $H$. 
The diagrams (g) and (h) shows the two-loop diagram which involves a mixture of two loop momenta, 
where the typical topologies of them are shown in Fig.~\ref{diag_hh2}. }
\label{diag_hh}
\end{center}
\end{figure}

\begin{figure}[t]
\begin{center}
\includegraphics[width=150mm]{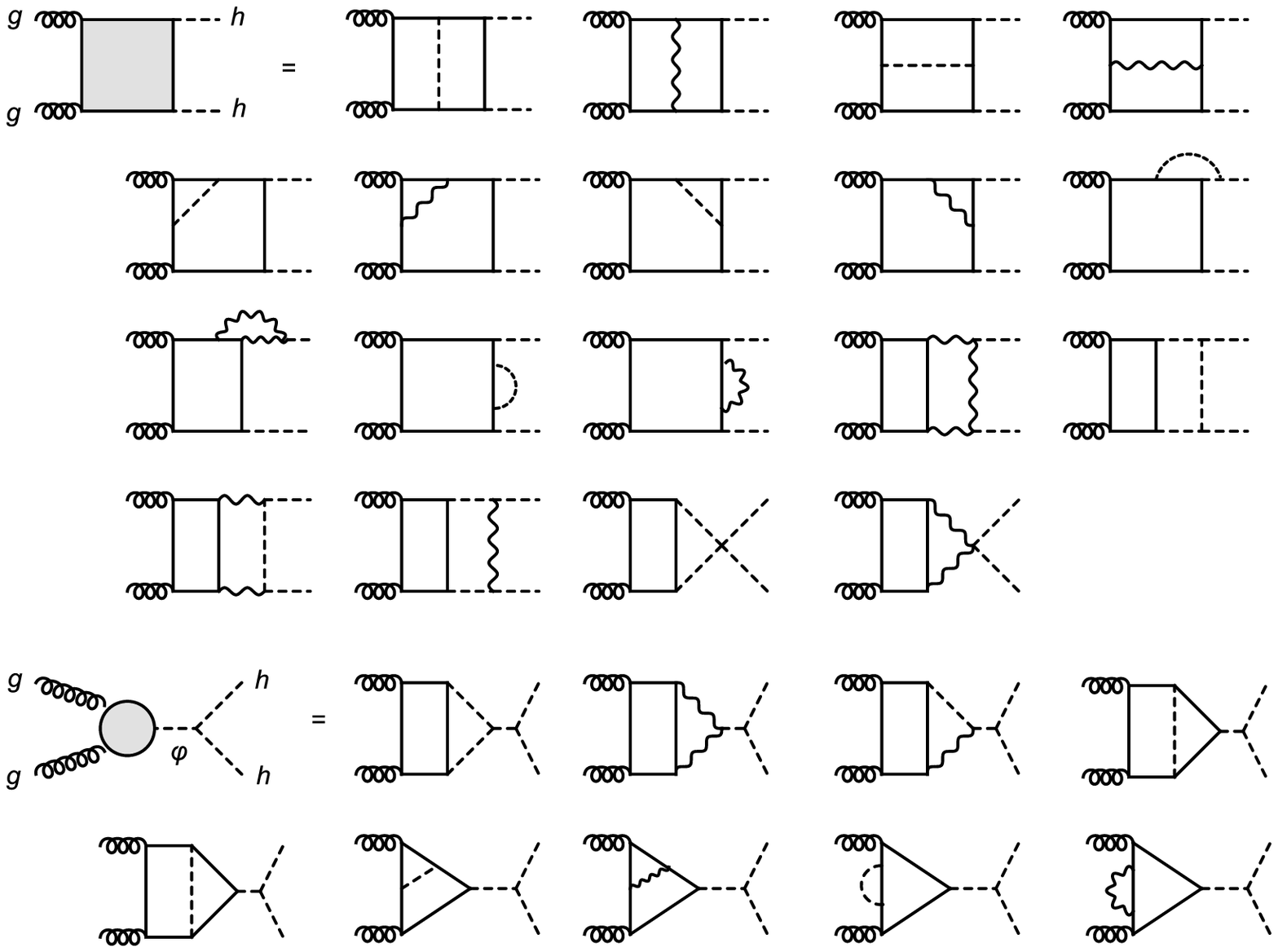}
\caption{
Typical topology of the two-loop diagrams (g) and (h) in Fig.~\ref{diag_hh}. 
}
\label{diag_hh2}
\end{center}
\end{figure}

We now ready to calculate the cross section of the double Higgs boson production via the gluon fusion process: $gg \to hh$ at the {\it partial} two-loop level, where 
the meaning of ``partial'' will be clarified below. 

The relevant Feynman diagrams are shown in Figs.~\ref{diag_hh} and \ref{diag_hh2}. 
Among the eight diagrams displayed in Fig.~\ref{diag_hh}, 
(a) and (b) correspond to the LO triangle and box type diagrams, respectively. 
All the other diagrams show the two-loop contributions.   
These two-loop diagrams can be separated into two categories, i.e., the diagrams (c)-(f) and those (g) and (h). 
The former one can be calculated by the product of the one-loop triangle or box diagram and one-loop corrections to the scalar three-point or two-point functions, 
where these two parts can be separately calculated with each other. 
On the other hand, the latter category has a mixture of two loop momenta, so that we need to evaluate the full two-loop integral. 
Another important difference between the former and latter contribution is found in the power of the scalar trilinear couplings $\lambda_{\varphi\varphi'\varphi''}$. 
The former (latter) contribution involves a cubic (quadratic) dependence on $\lambda_{\varphi\varphi'\varphi''}$. 
Therefore, when we take a 
large value of $\lambda_{\varphi\varphi'\varphi''}$ couplings within the extent allowed by the theoretical constraints, 
the deviation in the cross section of the double Higgs boson production mainly comes from the diagrams (a)-(e). 
On the other hand, if we take a small value of $\lambda_{\varphi\varphi'\varphi''}$, the contributions from (g) and (h) cannot be neglected. 
In this subsection, we only take into account the contributions from (a)-(e), and we call this level of the calculation as the partial two-loop level.
We note that the diagrams (e) and (f) vanish in our on-shell renormalization scheme explained in Sec.~IV when the on-shell Higgs boson $h$ is produced. 
Consequently, the diagrams (a)-(d) are taken into account in our calculation. 

It has been known that QCD corrections largely change the cross section of the $gg\to hh$ process. The NLO calculation in QCD has 
been evaluated in Ref.~\cite{DL}, and it has been clarified that the amount of the NLO correction is from $-30$ to +20\% level depending on the choice of $m_H$. 
In this paper, we calculate the cross section at LO in QCD.

In the HSM, the parton level cross section is calculated by 
\begin{align}
&\hat{\sigma} (gg \to hh)
 = \frac{G_F^2\alpha_s^2}{256(2\pi)^3} \int^{\hat{t}_{\text{max}}}_{\hat{t}_{\text{min}}} d\hat{t}
 \left(\Big|C_\Delta F_\Delta + c_\alpha^2 F_\Box\Big|^2  + \left|c_\alpha^2 G_\Box\right|^2 \right), \label{gghh1}
\end{align}
where $F_\Delta$ is the loop function for the triangle diagram, 
while $F_\Box$ $(G_\Box)$ is that for the box diagram with the same (opposite) helicity of the initial gluons. 
The analytic formulae for these loop functions are given in Ref.~\cite{hh}. 
In Eq.~(\ref{gghh1}),  $C_\Delta$ is the coefficient of the triangle diagram given as 
\begin{align}
C_\Delta &=  \sum_{\varphi = h,H}
c_\varphi\left[- \frac{v\hat{\Gamma}_{\varphi hh}(m_h^2,m_h^2,\hat{s})}{\hat{s}-m_\varphi^2 + im_\varphi \Gamma_\varphi}
+\frac{ v\Gamma_{\varphi hh}^{\text{tree}}}{(\hat{s}-m_\varphi^2+im_\varphi \Gamma_\varphi)^2}\hat{\Pi}_{\varphi\varphi}(\hat{s}) \right]\notag\\
&+\frac{v(c_\alpha \Gamma_{Hhh}^{\text{tree}} + s_\alpha \Gamma_{hhh}^{\text{tree}}) }
{(\hat{s}-m_H^2+im_H \Gamma_H)(\hat{s}-m_h^2+im_h \Gamma_h)}\hat{\Pi}_{Hh}(\hat{s}), 
\end{align}
where $c_\varphi=c_\alpha (s_\alpha)$ for $\varphi = h (H)$ and $\Gamma_\varphi$ is the width of $\varphi$. 
The total cross section in the $pp$ collision is calculated by convoluting the di-gluon parton luminosity function ${\cal L}_{gg}$:
\begin{align}
\sigma (pp \to gg \to hh) 
& = \int_{\tau_0}^1 d\tau \frac{d{\cal L}_{gg}}{d\tau} \hat{\sigma}(\hat{s} = \tau s), 
\end{align}
where $\tau_0 = 4m_h^2/s$ with $s$ being the collision energy of $pp$.

\begin{table}[t]
\begin{center}
\begin{tabular}{c||c|c|c|c}
\hline\hline & ($m_H$, $\sin\alpha$) & $\Delta\kappa_h$ [\%] & $\Gamma_H$ [GeV]& $\sigma_{\text{tot}}^{\text{HSM}}/\sigma_{\text{tot}}^{\text{SM}}$  \\  \hline
BP1    & $(200~\text{GeV},~0.1)$ & 66.1   & 1.41$\times 10^{-2}$  & 0.57             \\
BP2    & $(200~\text{GeV},~0.2)$ & 144   &  5.64$\times 10^{-2}$  & 0.44             \\
BP3    & $(400~\text{GeV},~0.1)$ & 32.6   &  0.580 & 5.32           \\
BP4    & $(400~\text{GeV},~0.2)$ & 105   &  2.22   & 20.0            \\
\hline\hline
\end{tabular} 
\end{center}
\caption{
BPs for the numerical evaluation of the cross section of $gg \to hh$.  
For all the BPs, we take $\lambda_{\Phi S}=1.9$, $\mu_S=\lambda_S=0$ and the collision energy $\sqrt{s}=13$ TeV. 
} \label{BPs}
\end{table}

\begin{figure}[t]
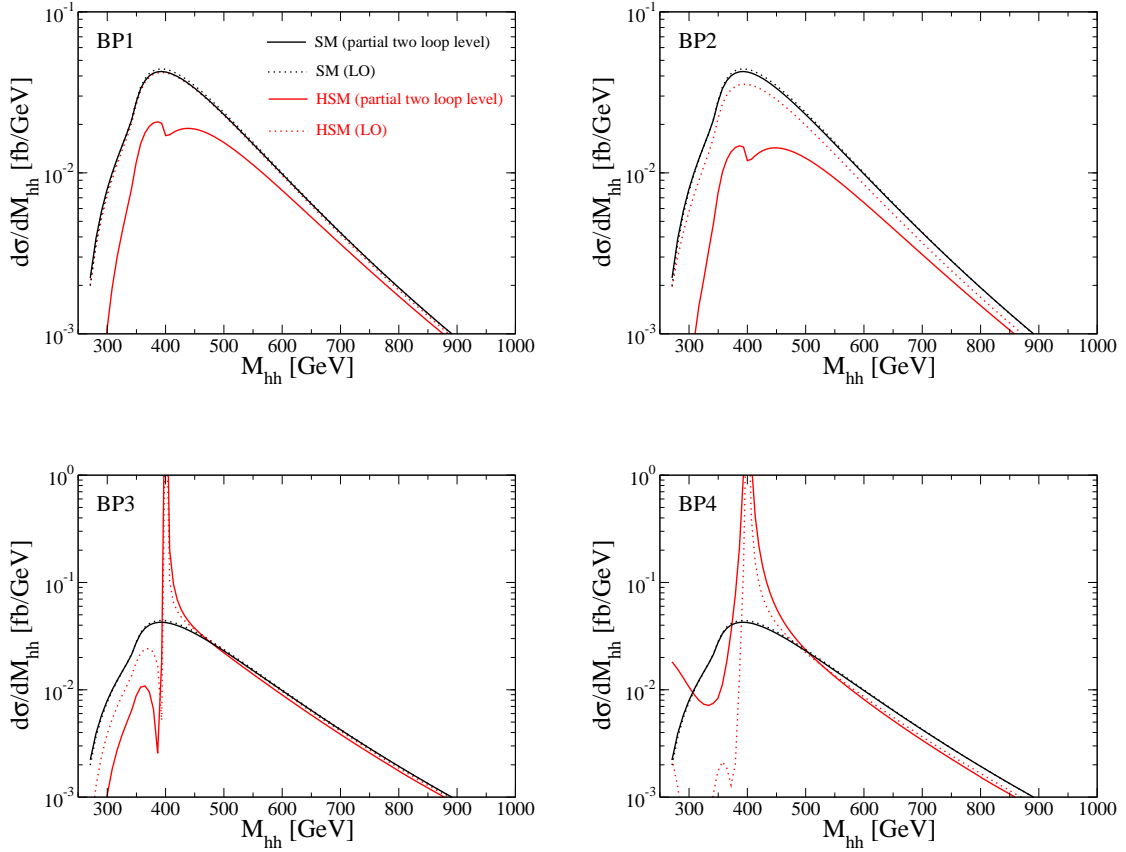

\begin{center}
\includegraphics[width=70mm]{dcross_BP1.eps} \hspace{5mm}
\includegraphics[width=70mm]{dcross_BP2.eps}\\ \vspace{10mm}
\includegraphics[width=70mm]{dcross_BP3.eps} \hspace{5mm}
\includegraphics[width=70mm]{dcross_BP4.eps}
\end{center}
\caption{
Differential hadronic cross sections of the double Higgs boson production process $gg \to hh$ in the $pp$ collision at $\sqrt{s}=13$ TeV. 
The upper-left, upper-right, lower-left and lower-right panels show the case for BP1, BP2, BP3 and BP4, respectively. 
The dotted (solid) curve shows the leading order (partial two-loop level) results, and the black (red) curve show the SM (HSM) results. 
}
\label{dcross}
\end{figure}

From now on, we present the numerical results of the double Higgs boson production cross section. 
For this analysis, we provide four benchmark points (BPs) as shown in Table~\ref{BPs}, in which 
we also give the outputs of $\Delta\kappa_h$, $\Gamma_H$ and the ratio of the total $gg\to hh$ cross section $\sigma_{\text{tot}}^{\text{HSM}}/\sigma_{\text{tot}}^{\text{SM}}$. 

In Fig.~\ref{dcross}, we show the differential hadronic cross section of the $gg \to hh$ process in the $pp$ collision at $\sqrt{s}=13$ TeV in the four BPs. 
In this calculation, we use the CTEQ6L~\cite{cteq} parton distribution functions, where its factorization scale $\mu_F$ is fixed to be $\sqrt{\hat{s}} = M_{hh}$ with $M_{hh}$ 
being the invariant mass of the $hh$ system. 
The black (red) curves shows the results in the SM (HSM), while the dotted (solid) curve shows the result at LO (the partial two-loop level). 
For BP1 and BP2 (upper two panels), the LO prediction in the HSM is almost the same as that in the SM. 
On the other hand, the prediction at the partial two-loop level is smaller than the corresponding SM result mainly due to 
larger distractive interference effects between the triangle and the box diagram contributions. 
In the red solid curves, we can observe the small dip at around $M_{hh}=400$ GeV ($=2\times m_H^{}$) which happens due to the threshold effect of $\hat{\Gamma}_{Hhh}$. 
For BP3 and BP4 (lower two panels), we see the significant difference between the results in the SM and in the HSM. 
In these cases, the peak at around $M_{hh} = 400$ GeV appears because of the resonance effect of the $H$ propagation, and its shape is quite narrow. 
This can be explained by the small width of $H$ as compared to the mass of $H$ as we saw in Fig.~\ref{width}. 
Thanks to this resonant effect of $H$, the total cross section significantly increases as compared to the case with $m_H<2m_h$ and the SM case, i.e., 
the ratio of the total cross section becomes 5.32 (20.0) in BP3 (BP4) as it is shown in Table~\ref{BPs}. 

\section{Conclusions}

We have calculated the one-loop correction to the triple scalar boson couplings $hhh$ and $Hhh$ based on the on-shell renormalization scheme in the HSM. 
We then applied these one-loop corrected couplings to calculate the decay rate of the $H\to hh$ mode and the double Higgs boson production process $gg\to hh$ at the LHC. 
It has been clarified that the one-loop correction to the $hhh$ coupling 
can change its tree level prediction to be the order of 100\% under the constraint from the perturbative unitarity, triviality, vacuum stability and conditions to avoid the wrong vacuum.  
We have found that 
the deviation in the $hhh$ coupling from the SM prediction can maximally be about 250\%, 150\% and 75\% for $m_H^{}=300$, 500 and 1000 GeV, respectively,  
under the requirement that the cutoff scale of the model is higher than 3 TeV. 
We have also shown the difference in the possible allowed value of the one-loop corrected $hhh$ coupling in the HSM and the Type-I THDM, namely,  
the decoupling behavior in these two models is quite different. 
The deviation in the $hhh$ coupling from the SM quickly reduces in the THDM, while the ${\cal O}(100)\%$ deviation can still remain in the HSM even at $m_H>1$ TeV. 
We have finally seen the cross section of the $gg \to hh$ process, where the cross section can significantly enhance due to the resonance effect of $H$. 
As an example when $m_{H}^{}=400$ GeV and $s_\alpha=0.1~(0.2)$, the cross section maximally becomes 5.32 (20.0) times larger than the SM prediction  at $\sqrt{s}=13$ TeV. 

\vspace*{4mm}
\noindent
\section*{Acknowledgments}
\noindent
The work of S.~K.~was supported in part by 
Grant-in-Aid for Scientific Research on Innovative Areas, The Ministry of Education, Culture, Sports, Science and Technology, No. 16H06492 and 
Grant H2020-MSCA-RISE-2014 no. 645722 (Non Minimal Higgs). 
K.~Y. was supported by a JSPS postdoctoral fellowships for research abroad. 

\vspace*{4mm}
\noindent
{\it Note added}--

\noindent
After this paper was completed, 
Ref.~\cite{He} appeared in which the one-loop correction to the $hhh$ coupling was calculated in the HSM.

\newpage
\begin{appendix}

\section{Scalar couplings}

The scalar trilinear and quartic couplings are defined by 
\begin{align}
{\cal L} = +\lambda_{\phi_1\phi_2\phi_3} \phi_1 \phi_2\phi_3 +
\lambda_{\phi_1\phi_2\phi_3\phi_4} \phi_1 \phi_2\phi_3\phi_4 + \cdots. 
\end{align}
These couplings are given by~\cite{HSM_KKY}
\begin{align}
\lambda_{hhh} &= -\frac{m_h^2}{2v}c_\alpha^3 -vc_\alpha s_\alpha^2 \lambda_{\Phi S} + s_\alpha^3 \mu_S , \\
\lambda_{HHh} &= -\frac{1}{2v}c_\alpha s_\alpha^2(m_h^2 + 2m_H^2) -\frac{v}{4}(c_\alpha+3c_{3\alpha}^{})\lambda_{\Phi S} 
+ 3c_\alpha^2s_\alpha \mu_S , \\
\lambda_{Hhh} &= -\frac{1}{2v}c_\alpha^2 s_\alpha(2m_h^2 + m_H^2) +\frac{v}{2}(1 +3c_{2\alpha}^{})s_\alpha \lambda_{\Phi S} 
- 3c_\alpha s_\alpha^2 \mu_S , \\
 \lambda_{HHH} &=
 - \frac{s_\alpha^3}{2v}m_H^2 
 - 4c_\alpha^3 \lambda_S^{} v_S^{} - c_\alpha^3 \mu_S^{}
 - s_\alpha^{} c_\alpha^{2} \lambda_{\Phi S}^{} v, \\
 \lambda_{G^0 G^0h} &= -\frac{m_h^2 c_\alpha}{2 v}, \\
 \lambda_{G^+ G^-h}&= -\frac{m_h^2 c_\alpha}{ v},  \\
 \lambda_{G^0 G^0H} &= -\frac{m_H^2 s_\alpha}{2 v}, \\
 \lambda_{G^+ G^-H} &=-\frac{m_H^2 s_\alpha}{v}, \\
 \lambda_{hhhh}^{} &=
 - (c_\alpha^2 m_h^2 + s_\alpha^2 m_H^2) \frac{c_\alpha^4}{8v^2}
 - s_\alpha^4 \lambda_S^{}
 - \frac{s_{2\alpha}^2}{8}\lambda_{\Phi S}^{}, \\
 \lambda_{Hhhh}^{} & =
 - \frac{c_\alpha^5 s_\alpha^{}}{2v^2} m_h^2 
 - \frac{s_{2\alpha}^3}{16v^2}m_H^2
 + 4 c_\alpha^{} s_{\alpha}^3 \lambda_S^{}
 + \frac{s_{4\alpha}^{}}{4} \lambda_{\Phi S}^{}, \\
 \lambda_{HHhh}^{} & =
 - (c_\alpha^2 m_h^2 + s_\alpha^{2} m_H^2)\frac{3s_\alpha^2 c_\alpha^2}{4v^2}
 - \frac{\lambda_{\Phi S}^{}}{8} (1+ 3c_{4\alpha}^{}) - 6\lambda_S^{} c_\alpha^2 s_\alpha^2, \\
 \lambda_{HHHh}^{} & =
 4\lambda_S^{}c_\alpha^3 s_\alpha^{}
 - \frac{m_H^2}{2v^2}c_\alpha s_\alpha^5
 - \frac{m_h^2}{16v^2}s_{2\alpha}^3 - \frac{\lambda_{\Phi S}^{}}{4}s_{4\alpha}^{},\\
 \lambda_{G^+G^-hh} &=
 - \frac{c_\alpha^4}{2v^2} m_h^2 - \frac{s_{2\alpha}^2}{8v^2} m_H^2
 - s_\alpha^2 \lambda_{\Phi S}^{}, \\
 \lambda_{G^+ G^-Hh} & =
 - (c_\alpha^2 m_h^2 + s_\alpha^2 m_H^2) \frac{s_\alpha^{}c_\alpha^{} }{v^2}
 + 2 s_\alpha^{} c_\alpha^{} \lambda_{\Phi S}^{}, \\
 \lambda_{G^+ G^-HH} &=
 - (4s_\alpha^4 m_H^2 + m_h^2 s_{2\alpha}^2) \frac{1}{8v^2}
 - c_\alpha^2 \lambda_{\Phi S}^{}, \\
 \lambda_{G^0G^0hh}^{} &=
 - \frac{m_h^2}{4v^2}c_\alpha^4 - \frac{m_H^2}{16v^2}s_{2\alpha}^2
 - \frac{\lambda_{\Phi S}}{2} s_\alpha^2, \\
 \lambda_{G^0G^0Hh} &=
 - \frac{m_h^2}{2v^2}c_\alpha^3 s_\alpha^{}
 - \frac{m_H^2}{2v^2}c_\alpha^{} s_\alpha^{3} + \lambda_{\Phi S}c_\alpha^{}s_\alpha^{}, \\
 \lambda_{G^0G^0HH} &=
 - \frac{m_h^2}{16v^2}s_{2\alpha}^2 -\frac{m_H^2}{4v^2}s_\alpha^4-\frac{1}{2}\lambda_{\Phi S}c_\alpha^2, \\
\lambda_{HHHH} & = -\frac{s_\alpha^4}{8v^2}(c_\alpha^2 m_h^2 + s_\alpha^2 m_H^2) -c_\alpha^4 \lambda_S -\frac{1}{8}s_{2\alpha}^2 \lambda_{\Phi S}. 
 \end{align}

\section{Beta functions}

We present the full set of the beta functions for the dimensionless couplings at the one-loop level. 
The beta functions for the gauge couplings and the top Yukawa coupling $y_t$ are the same form as those of the SM:
\begin{align}
\beta(g_3)&=\frac{g_3^3}{16\pi^2}(-7), \quad \beta(g_2)=\frac{g_2^3}{16\pi^2}\left(-\frac{19}{6}\right), \quad \beta(g_1)=\frac{g_1^3}{16\pi^2}\left(\frac{41}{6}\right), \\
\beta(y_t)&=\frac{1}{16\pi^2}\left[\frac{9}{2}y_t^3 -y_t\left(8g_3^2+\frac{9}{4}g_2^2+\frac{17}{12}g_1^2\right)\right]. 
\end{align}
Those for the dimensionless couplings in the potential are given~\cite{HSM-RGE} by 
\begin{align}
\beta(\lambda)&=\frac{1}{16\pi^2}\left[24\lambda^2 + 2\lambda_{\Phi S}^2
-6 y_t^4+\frac{9}{8}g_2^4+\frac{3}{8}g_1^{4}+\frac{3}{4}g_1^2g_2^{2}-\lambda(9g_2^2+3g_1^{2}-12y_t^2)\right], \\
\beta(\lambda_{\Phi S}^{})& =
\frac{1}{16\pi^2}
\left[12\lambda \lambda_{\Phi S}^{}+8\lambda_{\Phi S}^{2}+24\lambda_{\Phi S}^{} \lambda_S -\lambda_{\Phi S}^{}\left(\frac{9}{2}g_2^2  +\frac{3}{2}g_1^2 -6y_t^2\right)\right], \\
\beta(\lambda_{S}^{})& =\frac{1}{16\pi^2}\left(2\lambda_{\Phi S}^2 + 72\lambda_S^2 \right). \label{lams}
\end{align}

 \section{1PI diagrams}\label{sec:1PI}

We give the analytic expressions for the 1PI diagram contributions to the $hhh$ ($\Gamma_{hhh}^{\text{1PI}}$) and 
$Hhh$ ($\Gamma_{Hhh}^{\text{1PI}}$) vertices in terms of the Passarino-Veltman functions~\cite{Ref:PV}. 
The 1PI diagram contributions to the scalar one-point and two-point functions have been presented in Ref.~\cite{HSM_KKY}. 
In our calculation, we adopt the 't~Hooft--Feynman gauge, 
so that the masses of the Nambu-Goldstone bosons $m_{G^\pm}$ and $m_{G^0}$ 
and those of the Fadeev-Popov ghosts $m_{c^{\pm}}$, $m_{c^0}$ and $m_{c^\gamma}^{}$ are the same as corresponding masses of the gauge bosons.
For the Passarino-Veltman three point functions,  
we use the simplified form as $C_i[X, Y, Z] \equiv C_i[p_1^2, p_2^2, q^2; m_X, m_Y, m_Z]$, where $p_1^\mu$ and $p_2^\mu$ are incoming four-momenta and $q^\mu = p_1^\mu + p_2^\mu$. 

The fermion loop contributions to $\Gamma_{hhh}^{\text{1PI}}$ and $\Gamma_{Hhh}^{\text{1PI}}$ are given by 
 \begin{align}
  &(16\pi^2)\Gamma_{hhh}^{1\PI}(p_1^2,p_2^2,q^2)_F=
     -8c_\alpha^3\sum_fN_c^f \frac{m_f^4}{v^3}\Big[
   3(p_1^2C_{21} +p_2^2C_{22} +2p_1\cdot p_2 C_{23} +DC_{24}) 
   \notag\\
 & +2(2p_1^2+p_1\cdot p_2)C_{11}
   +2(2p_1\cdot p_2 +p_2^2)C_{12} 
   +(m_f^2+p_1^2+p_1\cdot p_2)C_0 \Big](f,f,f), \\
& \Gamma_{Hhh}^{1\PI}(p_1^2,p_2^2,q^2)_F= \Gamma_{hhh}^{1\PI}(p_1^2,p_2^2,q^2)_F\Big|_{c_\alpha^3 \to c_\alpha^2 s_\alpha}, 
 \end{align}
where $D = 4-2\epsilon$. 
The bosonic loop contributions are given by 
 \begin{align}
 & (16\pi^2)\Gamma_{hhh}^{1\PI}(p_1^2,p_2^2,q^2)_B=
     4\frac{m_W^4}{v^3}c_\alpha^3D\,  B_0[W,W]
   + 2\frac{m_Z^4}{v^3}c_\alpha^3D\, B_0[Z,Z]\notag\\[+5pt]
 & +2\lambda_{G^+G^-h}\lambda_{G^+G^-hh}  B_0[G^\pm,G^\pm] 
  +4\lambda_{G^0G^0h}\lambda_{G^0G^0hh}  B_0[G^0,G^0]  \notag\\[+5pt]
 & +72\lambda_{hhh}\lambda_{hhhh}B_0[h,h]
  +12\lambda_{Hhh}\lambda_{Hhhh}   B_0[h,H]  
   +4\lambda_{HHh}\lambda_{HHhh}   B_0[H,H] \notag\\[+5pt]
 &  +2 g^3 m_W^3 c_\alpha^3 D C_0[W,W,W]
   + g_Z^3 m_Z^3 c_\alpha^3 D C_0[Z,Z,Z] \notag\\[+5pt]
 & -\frac{1}{4}g_Z^3 m_Z c_\alpha^3 C_{VVS}^{hhh}[Z,Z,G^0] 
   -\frac{1}{2}g^3 m_W c_\alpha^3  C_{VVS}^{hhh}[W,W,G^\pm] \notag\\[+5pt] 
 &  +\frac{1}{2}g^2 c_\alpha^2 \lambda_{G^+G^-h}   ( C_{VSS}^{hhh}[W,G^\pm,G^\pm]+C_{SSV}^{hhh}[G^\pm,G^\pm,W]+C_{SVS}^{hhh}[G^\pm,W,G^\pm]   ) \notag\\
&   +\frac{1}{2}g^2 c_\alpha^2 \lambda_{G^0G^0h}    ( C_{VSS}^{hhh}[Z,G^0,G^0]+C_{SSV}^{hhh}[G^0,G^0,Z]+C_{SVS}^{hhh}[G^0,Z,G^0]  ) 
  \notag\\[+5pt] 
 & -2\lambda_{G^+G^-h}^3 C_0[G^\pm,G^\pm,G^\pm] 
   -8\lambda_{G^0G^0h}^3 C_0[G^0,G^0,G^0]  \notag\\[+5pt]
 & -216\lambda_{hhh}^3 C_0[h,h,h] 
   -24 \lambda_{Hhh}^2 \lambda_{hhh} \left\{
   C_0[h,h,H] +C_0[h,H,h] +C_0[H,h,h] \right\} \notag\\[+5pt]
 &  -8\lambda_{HHh}^3C_0[H,H,H]  
   -8 \lambda_{Hhh}^2 \lambda_{HHh} \left\{
   C_0[h,H,H] +C_0[H,h,H] +C_0[H,H,h] \right\} \notag\\[+5pt]
 & -\frac{1}{2} g^3 m_W^3 c_\alpha^3 
   C_0[c^\pm,c^\pm,c^\pm] 
   -\frac{1}{4} g_Z^3 m_Z^3 c_\alpha^3 
   C_0[c^0,c^0,c^0],
 \end{align}
 \begin{align}
 & (16\pi^2)\Gamma_{Hhh}^{1\PI}(p_1^2,p_2^2,q^2)_B=
     4\frac{m_W^4}{v^3}s_\alpha^{} c_\alpha^2D  B_0[W,W]
   + 2\frac{m_Z^4}{v^3}s_\alpha^{} c_\alpha^2D B_0[Z,Z]\notag\\[+5pt]
   & +2\lambda_{G^+G^-H}\lambda_{G^+G^-hh}  B_0[q^2; m_{G^\pm},m_{G^\pm}]
   + \lambda_{G^+G^-h}\lambda_{G^+G^-Hh}  (B_0[p_1^2; m_{G^\pm},m_{G^\pm}] + B_0[p_2^2; m_{G^\pm},m_{G^\pm}]) \notag\\[+5pt]
   &+4\lambda_{G^0G^0H}\lambda_{G^0G^0hh}  B_0[q^2;m_{G^0},m_{G^0}]
   +2\lambda_{G^0G^0h}\lambda_{G^0G^0Hh}  (B_0[p_1^2;m_{G^0},m_{G^0}]   +B_0[p_2^2;m_{G^0},m_{G^0}]  ) \notag\\[+5pt]
   & +24\lambda_{Hhh}\lambda_{hhhh}B_0[q^2;m_h,m_h] +
    18\lambda_{hhh}\lambda_{Hhhh} (B_0[p_1^2;m_h,m_h] +B_0[p_2^2;m_h,m_h] ) \notag\\[+5pt]
    & +12\lambda_{HHh}\lambda_{Hhhh}   B_0[q^2;m_h,m_H]
    + 8\lambda_{Hhh}\lambda_{HHhh}  ( B_0[p_1^2;m_h,m_H] + B_0[p_2^2;m_h,m_H]) \notag\\[+5pt]
    & +12\lambda_{HHH}\lambda_{HHhh}   B_0[q^2;m_H,m_H]
    +6\lambda_{HHh}\lambda_{HHHh}   (B_0[p_1^2;m_H,m_H] + B_0[p_2^2;m_H,m_H])\notag\\[+5pt]
 &  +2 g^3 m_W^3 s_\alpha^{}c_\alpha^2 D C_0[W,W,W]
   + g_Z^3 m_Z^3 s_\alpha^{}c_\alpha^2 D C_0[Z,Z,Z] \notag\\[+5pt]
 & -\frac{1}{4}g_Z^3 m_Z s_\alpha^{}c_\alpha^2 C_{VVS}^{hhh}[Z,Z,G^0] 
   -\frac{1}{2}g^3 m_W s_\alpha^{}c_\alpha^2 C_{VVS}^{hhh}[W,W,G^\pm] \notag\\[+5pt] 
   &  +\frac{g^2}{2} s_\alpha^{}c_\alpha^{} \lambda_{G^+G^-h} \{C_{VSS}^{hhh}
   [W,G^\pm,G^\pm] 
     + C_{SSV}^{hhh}[G^\pm,G^\pm,W]\} 
     +\frac{g^2}{2} c_\alpha^{2} \lambda_{G^+G^-H} C_{SVS}^{hhh}[G^\pm,W,G^\pm] \notag\\[+5pt] 
   &+\frac{g_Z^2}{2} s_\alpha^{}c_\alpha^{} \lambda_{G^0G^0h}\{C_{VSS}^{hhh}[Z,G^0,G^0] 
   + C_{SSV}^{hhh}[G^0,G^0,Z]\} 
   +\frac{g_Z^2}{2} c_\alpha^{2} \lambda_{G^0G^0H}
C_{SVS}^{hhh}[G^0,Z,G^0] 
  \notag\\[+5pt] 
 & -2\lambda_{G^+G^-h}^2\lambda_{G^+G^-H}^{} C_0[G^\pm,G^\pm,G^\pm] 
   -8\lambda_{G^0G^0h}^2\lambda_{G^0G^0H}^{} C_0[G^0,G^0,G^0]  \notag\\[+5pt]
 & -72\lambda_{hhh}^2\lambda_{Hhh}^{} C_0[h,h,h] 
   -24 \lambda_{HHh}^{}\lambda_{Hhh}^{} \lambda_{hhh}^{} \left\{
   C_0[h,h,H] +C_0[H,h,h] \right\} - 8\lambda_{Hhh}^3C_0[h,H,h]\notag\\[+5pt]
 &  -24\lambda_{HHh}^2\lambda_{HHH}^{} C_0[H,H,H]  
   -8 \lambda_{HHh}^2 \lambda_{Hhh} \left\{
   C_0[h,H,H] +C_0[H,H,h] \right\}  - 24\lambda_{HHH}\lambda_{Hhh}^2C_0[H,h,H]\notag\\[+5pt]
 & -\frac{1}{2} g^3 m_W^3 s_\alpha^{}c_\alpha^2 
   C_0[c^\pm,c^\pm,c^\pm] 
   -\frac{1}{4} g_Z^3 m_Z^3 s_\alpha^{}c_\alpha^2 
   C_0[c^0,c^0,c^0],
 \end{align}
where we define the following functions;
 \begin{align}
 B_0[X,Y] &= B_0[q^2;m_X^{},m_Y^{}] +B_0[p_1^2;m_X^{},m_Y^{}] +B_0[p_2^2;m_X^{},m_Y^{}], \\
 C_{VVS}^{hhh}[X,Y,Z]&= \{p_1^2(C_{21}+3C_{11} +2C_0) +p_2^2(C_{22}+4C_{12} +4C_0)
    \notag\\[+5pt] 
 & +p_1\cdot p_2 (2C_{23}+4C_{11} +6C_0 +3C_{12}) +DC_{24}\}[X,Y,Z] \notag\\[+5pt]
 & + \{p_1^2(C_{21}+3C_{11} +2C_0) +p_2^2(C_{22}-C_{12} )
    \notag\\[+5pt] 
 & +p_1\cdot p_2 (2C_{23}-C_{11} -2C_0 +3C_{12}) +DC_{24}\}[X,Z,Y] \notag\\[+5pt]
 &+\{ p_1^2(C_{21} -2C_{11} +C_0) +p_2^2(C_{22}-C_{12})
    \notag\\[+5pt] 
 & +p_1\cdot p_2 (2C_{23} -C_{11} +C_0 -2C_{12}) +DC_{24}\}[Z,Y,X],
 \end{align}
 \begin{align}
  C_{VSS}^{hhh}[X,Y,Z]&=\{ p_1^2(C_{21}+4C_{11} +4C_0) +p_2^2(2C_{12}+C_{22})  \notag\\ 
 & +2p_1\cdot p_2 (C_{23}+C_{11} +2C_0 +2C_{12}) +DC_{24} \}[X,Y,Z] , \\
  C_{SSV}^{hhh}[X,Y,Z]&=
  \{p_1^2(C_{21} -C_0) +p_2^2(C_{22}-2C_{12} +C_0) \notag\\
   &+2p_1\cdot p_2 (C_{23}-C_{11} ) +DC_{24}\}[X,Y,Z], \\
  C_{SVS}^{hhh}[X,Y,Z]&=
 \{p_1^2(C_{21}-C_0) +p_2^2(C_{22}+2C_{12}) \notag\\[+5pt]
 &  +2p_1\cdot p_2 (C_{23}+C_{11} - C_0 ) +DC_{24}\}[X,Y,Z]. 
 \end{align}

\end{appendix}

\end{document}